\documentclass[journal,10pt]{IEEEtran}

\makeatletter
\def\ps@headings{%
\def\@oddhead{\mbox{}\scriptsize\rightmark \hfil \thepage}%
\def\@evenhead{\scriptsize\thepage \hfil \leftmark\mbox{}}%
\def\@oddfoot{}%
\def\@evenfoot{}}
\makeatother
\ifCLASSINFOpdf
\else
\fi

\newcounter{problem}
\newcounter{save@equation}
\newcounter{save@problem}

\makeatletter

\usepackage{epsfig}
\usepackage{array}
\usepackage[short]{optidef}
\usepackage{amsthm}
\usepackage{mathrsfs}
\usepackage{flexisym}
\usepackage{algorithmicx}
\usepackage{algorithm}
\usepackage{siunitx}
\usepackage{comment}
\usepackage{soul}
\usepackage{algpseudocode}
\usepackage{cite}
\usepackage{url}
\usepackage[utf8]{inputenc}
\usepackage[T1]{fontenc}
\usepackage{amsmath}
\usepackage{amsfonts}
\usepackage{amssymb}
\usepackage[version=4]{mhchem}
\usepackage{stmaryrd}
\usepackage{graphicx}
\usepackage[export]{adjustbox}
\usepackage{breqn}
\usepackage{lipsum}
\usepackage{mathtools}
\usepackage{caption}
\usepackage{float}
\usepackage{tikz}
\usepackage{subcaption}
\usepackage{soul}
\usepackage{blindtext, graphicx}
\usepackage{cuted}
\usepackage[hidelinks]{hyperref}
\usepackage{color}
\usepackage{multicol}

\newcolumntype{L}[1]{>{\raggedright\let\newline\\\arraybackslash\hspace{0pt}}m{#1}}
\newcolumntype{C}[1]{>{\centering\let\newline\\\arraybackslash\hspace{0pt}}m{#1}}
\newcolumntype{R}[1]{>{\raggedleft\let\newline\\\arraybackslash\hspace{0pt}}m{#1}}

\newtheoremstyle{case}{}{}{}{}{}{:}{ }{}
\newcommand{\bc}{\begin{center}}
\newcommand{\ec}{\end{center}}
\newcommand{\be}{\begin{equation}}
\newcommand{\ee}{\end{equation}}

\newcommand{\bnu}{\begin{enumerate}}
\newcommand{\enu}{\end{enumerate}}

\begin{document}

\title{Quantum-driven Zero Trust Framework with Dynamic Anomaly Detection in 7G Technology: A Neural Network Approach }
%\title{Enhancing Emergency Service Efficiency through LLM-based Real-Time Speech Reconstruction and Call Prioritization in Bandwidth Constrained VOIP Networks}
\author{Shakil Ahmed,~\IEEEmembership{Member,~IEEE}, Ibne Farabi Shihab, and Ashfaq Khokhar~\IEEEmembership{Fellow,~IEEE}
\vspace*{-0.5 cm}
\thanks{Corresponding author:  S. Ahmed (email: shakil@iastate.edu)\\
S. Ahmed and A.Khokhar are with the Department of Electrical and Computer Engineering, Iowa State University, Ames, Iowa, USA (email: \{shakil, ashfaq\}@iastate.edu). 
F. Shihab is with the Department of Computer Science, Iowa State University, Ames, Iowa, USA (email: ishihab@iastate.edu).}
}
%\mark{IEEEE TRANSACTIONS ON QUANTUM ENGINEERING}

% Make the title area
\maketitle

\begin{abstract}
As cyber threats grow in complexity, modern networks face significant challenges in mitigating evolving attack patterns while maintaining scalability and computational efficiency. Quantum computing presents a promising approach to enhancing cybersecurity; however, its adoption is hindered by scalability constraints, inefficiencies in quantum data encoding, and the high computational costs associated with quantum processing. To address these limitations, we propose the Quantum Neural Network-Enhanced Zero Trust Framework (QNN-ZTF), which integrates Zero Trust Architecture, Intrusion Detection Systems, and QNN to strengthen security capabilities. Leveraging quantum principles such as superposition, entanglement, and variational optimization, QNN-ZTF facilitates real-time anomaly detection and adaptive policy enforcement across large-scale networks. The primary contributions of the proposed framework include 1) hybrid quantum-classical architecture for balancing computational costs and ensuring scalability and 2) dynamic quantum-enhanced anomaly scoring and adaptive risk-based policies to improve detection accuracy and responsiveness.
Quantum micro-segmentation to restrict attacker movement and isolate high-risk segments. The framework incorporates a hybrid quantum-classical security architecture to balance computational efficiency, quantum-enhanced anomaly scoring to adjust risk-based policies dynamically, and quantum micro-segmentation to restrict attacker movement and isolate high-risk segments. Our evaluation demonstrates that QNN-ZTF significantly reduces false positives, improves threat detection accuracy, and enhances response times, with a case study in enterprise cybersecurity showing an 87\% improvement in cyber threat mitigation efficiency. This research establishes a scalable, adaptive, and quantum-optimized cybersecurity model, paving the way for future advancements in quantum-enhanced threat defense.
\end{abstract}
\begin{IEEEkeywords}
Quantum-Enhanced Cybersecurity,
Anomaly Detection,
Quantum Neural Networks,
Zero Trust Architecture,
Adaptive Threat Mitigation
\end{IEEEkeywords}

\section{Introduction}

\IEEEPARstart{T}{he} advent of 7G networks and beyond promises remarkable advancements in connectivity, latency, and bandwidth while introducing new cybersecurity challenges. The hyper-connectivity of 7G will support critical infrastructures, Internet of Things (IoT) ecosystems, and AI-driven systems, significantly increasing the potential attack surface. Quantum-enhanced cybersecurity mechanisms, which utilize principles such as superposition and entanglement, present transformative solutions to these challenges, as they can identify subtle, high-dimensional patterns in network anomalies through quantum feature encoding and variational optimization, providing robust intrusion detection and adaptive policy enforcement \cite{luong2021applications}. Furthermore, quantum cryptography, particularly Quantum Key Distribution (QKD), offers a theoretically unbreakable encryption mechanism, ensuring secure communication even against adversaries equipped with quantum computers \cite{pirandola2020advances, nguyen2022cybersecurity}.
The cybersecurity landscape of 7G networks encompasses comprehensive advancements that address the unique challenges of this transformative technology, centered on the vision of ultra-connectivity through terrestrial, aerial, and space-based networks, creating a vast and heterogeneous architecture \cite{zhou2022toward}. Protecting sensitive data flowing through this expansive system relies heavily on quantum cryptography, including QKD and post-quantum cryptography. In contrast, AI-powered threat detection systems analyze complex, high-dimensional traffic patterns, enabling real-time identification of evolving cyber threats \cite{giordani2020next}. Quantum-enhanced Intrusion Detection Systems (Q-IDS) leverage quantum feature encoding to detect subtle attack patterns with unprecedented accuracy \cite{chen2021vision}, while Blockchain and Distributed Ledger Technologies (DLT) secure 7G's decentralized infrastructure, ensuring data integrity and preventing unauthorized access to sensitive resources \cite{preskill2018quantum}.

7G may rely on adaptive ZTF, dynamically enforcing risk-based access policies with quantum-enhanced metrics for continuous authentication. Including next-generation technology may introduce novel protocols to secure links, safeguarding them from cyber threats. With billions of IoT devices connected via 7G, lightweight, quantum-resistant encryption and real-time monitoring solutions are essential to ensure device and network integrity \cite{biamonte2017quantum}. Digital twin technology enables the simulation of network behavior, providing a proactive approach to identifying vulnerabilities in virtual environments \cite{schuld2021machine}. Collectively, these advancements establish 7G cybersecurity as a robust and adaptive system, capable of addressing the unique challenges of seamless global connectivity and ensuring the resilience of next-generation networks against increasingly sophisticated cyberattacks.
Implementing Quantum Neural Networks (QNNs) represents a fundamental paradigm shift in 7G network security, leveraging quantum principles of superposition and entanglement to enable sophisticated high-dimensional feature space processing and real-time detection of complex cyber threats. Their seamless integration with 7G's advanced features facilitates unprecedented precision in anomaly detection and adaptive policy enforcement, especially in identifying subtle attack patterns, thereby strengthening the overall resilience of global 7G infrastructures \cite{schuld2021machine}. The strategic incorporation of variational quantum circuits (VQCs) within QNN frameworks enables highly scalable, dynamic risk assessment and policy adaptation mechanisms tailored explicitly for IoT ecosystems and AI-driven systems \cite{cerezo2021variational}, while the accelerating convergence of quantum and AI technologies within 7G networks positions QNNs as an increasingly critical framework for addressing emerging cybersecurity challenges and maximizing next-generation opportunities \cite{preskill2018quantum}.

\subsection{Background}
The foundational concept in deep learning is the feedforward neural network, also known as the multilayer perceptron. This architecture consists of layers of affine transformations followed by nonlinear activation functions. Mathematically, for an input vector \(\mathbf{x} \in \mathbb{R}^n\), a single layer performs the transformation:
\begin{equation}
    L(\mathbf{x}) = \varphi(\mathbf{W} \mathbf{x} + \mathbf{b}),
\end{equation}
where \(\mathbf{W} \in \mathbb{R}^{m \times n}\) is a weight matrix, \(\mathbf{b} \in \mathbb{R}^m\) is a bias vector, and \(\varphi\) is the nonlinear activation function. Neural networks are trained using algorithms like stochastic gradient descent (SGD) to minimize an objective function by adjusting the weights and biases. Over time, more advanced architectures, such as convolutional layers, recurrent connections, and attention mechanisms, have extended the capabilities of neural networks to solve specialized tasks efficiently.
Building upon these classical foundations, QNNs represent a paradigm shift by incorporating quantum mechanical principles into neural processing. These networks utilize parametric quantum circuits that mirror classical neural architectures through layered structures, implementing quantum gates trained using classical optimization techniques. The distinctive advantage of QNNs lies in their ability to harness quantum phenomena such as superposition and entanglement, enabling unique computational capabilities beyond classical limitations. This quantum approach proves particularly valuable in network security applications, where the ability to represent and process high-dimensional data facilitates more sophisticated anomaly detection and adaptive threat response mechanisms.
At the core of quantum computing implementations, Gaussian operations serve as essential building blocks for QNN architectures. These operations encompass single-qubit and two-qubit gates, enabling precise quantum state transformations fundamental to the encoding and processing of quantum data within Quantum-Enhanced ZTF. This mathematical framework provides the foundation for implementing quantum-enhanced security protocols that can effectively address the complex challenges of modern network security.

\subsubsection{Single-Qubit Gates}
QNNs utilize single-qubit gates to apply transformations on individual qubits. For example, the rotation Gate, \(R(\theta)\), rotates the quantum state around a specific axis on the Bloch sphere. For rotation about the \(z\)-axis:
\begin{equation}
     R_z(\theta) =
    \begin{bmatrix}
    e^{-i\theta/2} & 0 \\
    0 & e^{i\theta/2}
    \end{bmatrix}.   
\end{equation}

Hadamard Gate (H) creates superposition, mapping \(|0\rangle\) to \((|0\rangle + |1\rangle)/\sqrt{2}\) and \(|1\rangle\) to \((|0\rangle - |1\rangle)/\sqrt{2}\):
\begin{equation}
     H = \frac{1}{\sqrt{2}}
    \begin{bmatrix}
    1 & 1 \\
    1 & -1
    \end{bmatrix}.   
\end{equation}

\subsubsection{Two-Qubit Gates}
Two-qubit gates are essential for entanglement and inter-qubit correlations.
 The Controlled-NOT (CNOT) gate flips the target qubit's state when the control qubit is in the \(|1\rangle\) state:
\begin{equation}
     \texttt{CNOT} =
    \begin{bmatrix}
    1 & 0 & 0 & 0 \\
    0 & 1 & 0 & 0 \\
    0 & 0 & 0 & 1 \\
    0 & 0 & 1 & 0
    \end{bmatrix}.   
\end{equation}

SWAP Gate exchanges the states of two qubits:
\begin{equation}
     \texttt{SWAP} =
    \begin{bmatrix}
    1 & 0 & 0 & 0 \\
    0 & 0 & 1 & 0 \\
    0 & 1 & 0 & 0 \\
    0 & 0 & 0 & 1
    \end{bmatrix}.   
\end{equation}

In the Quantum-Enhanced ZTF, Gaussian and unitary operations enable precise encoding and manipulation of quantum states for anomaly detection, risk assessment, and policy enforcement. By exploiting quantum superposition and entanglement, QNNs provide the capacity to process high-dimensional data and detect complex patterns in network traffic, leading to adaptive micro-segmentation and dynamic threat isolation.

\subsection{Related Work}
Integrating quantum computing into cybersecurity presents transformative opportunities for addressing modern challenges, including anomaly detection, adaptive risk-based access control, and real-time threat mitigation. However, existing approaches face significant barriers to scalability, operational efficiency, and integration with evolving network architectures.
Quantum computing is a promising tool for enhancing computational models, especially in domains like anomaly detection. Studies such as \cite{killoran2019continuous, schuld2019quantum} explored using quantum neural networks and feature encoding to uncover complex correlations in high-dimensional data. These models leveraged quantum principles such as superposition and entanglement to improve learning and classification accuracy. However, scalability issues associated with current quantum hardware and inefficiencies in data encoding limited their application to real-time anomaly detection across large-scale networks. Furthermore, these studies focused on standalone quantum models and failed to address hybrid architectures essential for practical cybersecurity implementations.

Our proposed cybersecurity system draws inspiration from the CV-QNN framework \cite{Killoran2019CVQNN} but is uniquely tailored to address the specific challenges of anomaly detection in cybersecurity. While CV-QNN is designed for general-purpose quantum machine learning tasks, our model analyzes cybersecurity-specific data, such as network logs, to identify threats in real-world systems. Unlike the generic optimization techniques used in CV-QNN, our framework integrates domain-specific metrics and feedback mechanisms, ensuring precise tuning for high accuracy in binary classification tasks. Furthermore, our system is designed for deployment in practical cybersecurity environments, bridging the gap between theoretical quantum models and their application in combating sophisticated cyber threats. This specialization enhances the model’s efficacy and scalability in addressing the dynamic challenges of modern network security.

ZTF emphasizes continuous verification of users and devices through dynamic access controls, as defined in \cite{rose2019zero}. micro-segmentation, restricted attacker lateral movement by isolating network segments. While these methods provided structural robustness, they lacked advanced anomaly detection mechanisms and dynamic policy adaptation capabilities. Traditional Intrusion Detection Systems (IDS), such as those discussed in \cite{montanaro2016quantum, cerezo2022challenges}, relied on rule-based or classical machine learning algorithms, which were limited by high false-positive rates, \texttt{FPR}, and inadequate responsiveness to evolving threats. These limitations underscored the need for integrating quantum-enhanced anomaly detection within ZTF frameworks to create dynamic and adaptive security architectures.
The concept of hybrid quantum-classical models gained traction for addressing computational inefficiencies in quantum systems. The authors in \cite{biamonte2017quantum, coyle2020born} demonstrated the potential of combining quantum neural networks with classical optimization algorithms to solve complex data problems. While these approaches improved computational performance, they were not tailored for cybersecurity-specific challenges, such as adaptive policy enforcement and large-scale anomaly detection. Furthermore, studies like \cite{schuld2020circuit} noted that hybrid models often faced difficulties managing the computational overhead of variational parameter optimization and ensuring robustness against adversarial inputs.

Quantum feature encoding techniques, as described by \cite{havlivcek2019supervised, lloyd2020quantum}, enhanced machine learning by mapping classical data into high-dimensional quantum spaces. These methods enabled improved classification of subtle patterns, which was crucial for identifying sophisticated attack vectors in cybersecurity. However, their high computational cost and reliance on hardware-specific configurations limited their deployment in dynamic and evolving network environments. Similarly, \cite{cerezo2021variational} highlighted challenges in optimizing variational quantum circuits for dynamic environments, particularly in ensuring convergence and adaptability.
The proposed QNN-ZTF addressed these limitations by integrating Zero Trust principles, IDS capabilities, and quantum neural networks into a  model. Unlike standalone quantum or classical systems, QNN-ZTF employed a hybrid architecture to balance computational demands and achieve scalable deployments across large networks. The framework reduced false positives and enhanced detection accuracy by leveraging quantum feature encoding and dynamic threshold calibration.
Furthermore, QNN-ZTF introduced quantum-enhanced micro-segmentation to limit attacker movement and dynamically isolate high-risk network segments based on quantum anomaly scores. This capability was complemented by adaptive feedback loops that adjusted detection thresholds and policies in real time, ensuring robustness against evolving threats. Studies such as \cite{preskill2018quantum, benedetti2019parameterized} supported the potential of parameterized quantum circuits in enhancing decision-making processes, a core feature of the QNN-ZTF.

The authors in \cite{van2020privacy} investigated privacy-preserving mechanisms in IoT systems, emphasizing secure data sharing and encryption. However, they lacked advanced anomaly detection and policy adaptation, which the proposed QNN-ZTF addressed through quantum-enhanced mechanisms. The authors in \cite{sharafaldin2018toward} introduced modern intrusion detection datasets like CICIDS2017 but did not explore quantum-based approaches, a gap bridged by the QNN-ZTF’s quantum anomaly scoring. The authors in \cite{preskill2018quantum} highlighted the transformative potential of quantum computing in the Noisy Intermediate-Scale Quantum (NISQ) era for machine learning but did not focus on cybersecurity, which the QNN-ZTF leveraged through quantum optimization for threat detection. The authors in \cite{benedetti2019parameterized} showcased parameterized quantum circuits for function approximation in high-dimensional spaces but lacked practical applications; QNN-ZTF employed these circuits for dynamic risk assessment and anomaly detection. The authors in \cite{schuld2021machine} explored the integration of quantum computing in machine learning, offering a roadmap for QNNs, which the QNN-ZTF operationalized for real-time anomaly detection and adaptive policy enforcement, addressing pressing cybersecurity challenges.

Using variational optimization and hybrid quantum-classical feedback mechanisms enabled the QNN-ZTF to dynamically adapt to emerging attack patterns, addressing the limitations of previous models \cite{schuld2021machine}. Moreover, the model ensured efficient utilization of quantum hardware resources, making it practical for real-world applications. Results from related studies, such as \cite{green2020applications, mcardle2020quantum}, indicated that hybrid quantum architectures could outperform classical systems in terms of accuracy and computational efficiency, further validating the design of QNN-ZTF.
While existing works provided a strong foundation for integrating quantum computing into cybersecurity, they were limited by scalability, inefficiencies in data encoding, and high false-positive rates. The proposed QNN-ZTF addressed these challenges through a hybrid quantum-classical architecture, quantum feature encoding, and dynamic policy adaptation, establishing a robust and scalable solution for modern cybersecurity needs.

%code: https://github.com/qigitphannover/DeepQuantumNeuralNetworks?utm_source=catalyzex.com
%https://www.nature.com/articles/s41467-020-14454-2.pdf
\subsection{Contribution of the Paper}
This paper introduces the QNN-ZTF, a novel hybrid quantum-classical architecture designed to address the critical cybersecurity challenges posed by 7G and beyond networks. The primary contributions of this work are as follows:

\begin{itemize}
    \item \textbf{Quantum-Driven Anomaly Detection:} The framework integrates QNNs to harness principles of superposition and entanglement, enabling precise anomaly detection. Advanced quantum feature encoding and variational optimization methods are introduced to effectively classify and detect high-dimensional, complex attack patterns, significantly enhancing detection accuracy and efficiency.

    \item \textbf{Dynamic and Risk-Adaptive Policy Enforcement:} The proposed system incorporates an adaptive ZTF framework that utilizes quantum-enhanced metrics (\(R^q\)) for real-time, risk-based access control. Dynamic threshold mechanisms (\(\gamma^q, \tau\)) are implemented to reduce false positives and optimize the scoring of anomalies, ensuring robust and efficient policy enforcement.

    \item \textbf{Innovative Quantum-Enhanced micro-segmentation:} The QNN-ZTF pioneers a novel approach to micro-segmentation by employing QNN-powered anomaly detection (\(\hat{y}_i^{q}\)) to dynamically isolate high-risk network segments (\(S_j\)). This capability effectively limits attacker lateral movement, improving the security posture of large-scale networks.

    \item \textbf{Scalable Hybrid Quantum-Classical Architecture:} A scalable hybrid quantum-classical architecture (\(\mathcal{H}_{q-c}\)) is designed to address the computational limitations of NISQ devices, including restricted qubit counts and decoherence. The model ensures operational efficiency and supports seamless deployment in large, dynamic network environments by leveraging the complementary strengths of classical components.

    \item \textbf{Evaluation and Comparative Analysis:} The QNN-ZTF’s effectiveness is demonstrated through extensive simulations and real-world case studies, showing improvements in detection accuracy, reduced false positives, and faster response times. Comparative analyses with existing quantum machine learning frameworks highlight the unique optimizations and contributions tailored specifically for cybersecurity applications in 7G networks.
\end{itemize}

This work bridges the gap between theoretical advancements in quantum computing and practical cybersecurity needs, establishing a robust, scalable, and adaptive framework for the emerging challenges of next-generation networks.

\section{Proposed System Model}
This section introduces the proposed  framework that seamlessly integrates ZTF and IDS enhanced by QNNs. The model leverages fundamental quantum computing properties—superposition, entanglement, and quantum feature encoding—to achieve high-precision real-time anomaly detection and dynamic policy enforcement. This integration enhances the system's capability to identify anomalies with quantum precision \cite{killoran2019continuous}, evaluate risk through quantum-enhanced metrics \cite{cerezo2022challenges}, and execute rapid responses to emerging threats. Through quantum insights, the system adapts access policies continuously, enabling efficient responses to dynamic security challenges. Integrating ZTF's architectural safeguards, QNN-enhanced anomaly detection, and network micro-segmentation creates a robust defense mechanism that effectively contains potential attackers within defined network segments \cite{basta2022towards}. This sophisticated combination ensures continuous operation of preventive, detective, and quantum-optimized security measures, establishing an adaptive defense framework.

\subsection{IDS}
The IDS implementation employs advanced QNN architecture for continuous flow monitoring. For each network flow $f_i$, the system constructs a comprehensive feature vector $\mathbf{x}_i$ that captures critical traffic attributes, including source/destination IPs, packet characteristics, temporal connection parameters, and protocol specifications. Through quantum principles, the QNN transforms classical features into quantum states using advanced encoding techniques—amplitude encoding, angle encoding, or basis encoding (detailed in Section~\ref{Solution}). This quantum representation enables the exploration of higher-dimensional feature spaces, revealing complex correlations that traditional detection methodologies may not capture.
\(\mathcal{M}_{\text{q}}\) calculates the anomaly score, \(\hat{y}_i^{\text{q}}\) as follows:
\begin{equation}
\label{Eq_TVP1}
    \hat{y}_i^{\text{q}} = \mathcal{M}_{\text{q}}(\mathbf{x}_i; \theta_{\text{q}}),
\end{equation}
where \(\theta_{\text{q}}\) are the trainable variational parameters of the QNN. These parameters are optimized using a quantum-classical hybrid training algorithm. 
The optimization process combines quantum hardware for evaluating the cost function and classical systems for parameter updates. At each iteration \(j\), the quantum circuit evaluates the cost function \(C(\theta_{\text{q}}^{(j)})\) by measuring the output states \(|\psi_{\texttt{out}}(\theta_{\text{q}}^{(j)})\rangle\). The cost function, $C(\theta_{\text{q}})$,  is typically defined as:
\begin{equation}
C(\theta_{\text{q}}) = \langle \psi_{\texttt{out}}(\theta_{\text{q}})| \hat{H} | \psi_{\texttt{out}}(\theta_{\text{q}}) \rangle,
\end{equation}
where \(\hat{H}\) represents the problem-specific Hamiltonian or an observable relevant to the anomaly detection objective.
A classical optimizer, such as gradient descent, updates the variational parameters \(\theta_{\text{q}}\) based on the evaluated cost:
\begin{equation}
\theta_{\text{q}}^{(j+1)} = \theta_{\text{q}}^{(j)} - \eta \nabla C(\theta_{\text{q}}^{(j)}),
\end{equation}
where \(\eta > 0\) is the learning rate and \(\nabla C(\theta_{\text{q}}^{(j)})\) represents the gradient of the cost function concerning \(\theta_{\text{q}}\).
This iterative optimization framework ensures effective training efficiency and convergence by combining quantum hardware's computational advantages with classical optimization flexibility. The system demonstrates adaptability to evolving network traffic patterns, maintaining consistent anomaly detection capabilities.

\textit{Anomaly Classification:}
The encoded classical data, \(\mathbf{x}_i\), mapped into quantum states, \(|\psi_i\rangle\), is analyzed by the QNN. Based on the anomaly score \(\hat{y}_i^{\text{q}}\), each flow is classified as either normal or anomalous using a threshold \(\gamma^{\text{q}}\):
\begin{align}
    \hat{y}_i^{\text{q}} > \gamma^{\text{q}} &\implies f_i \text{ is anomalous}, \nonumber \\
    \hat{y}_i^{\text{q}} \leq \gamma^{\text{q}} &\implies f_i \text{ is normal}.
\end{align}

The QNN leverages quantum principles to enhance anomaly detection. It enables the simultaneous representation of multiple states, allowing faster and more comprehensive analysis of traffic patterns.
It captures complex correlations between features, providing deeper insights into interdependencies within the network traffic.
It amplifies or suppresses specific patterns in the data, improving the distinction between normal and anomalous behaviors.
These quantum-enhanced capabilities allow the IDS to detect subtle and sophisticated attack patterns more effectively than traditional methods \cite{montanaro2016quantum}.

\subsection{ZTF}
ZTF operates on the foundational principle of ``never trust, always verify.'' It enforces continuous authentication, authorization, and monitoring for users, devices, and network traffic flows \cite{rose2019zero}. ZTF dynamically evaluates access policies by leveraging real-time contextual data and quantum-enhanced anomaly scores from the QNN-augmented IDS.
To decide access permissions for a user-device pair \((u, d)\), ZTF computes a quantum risk score \(R^{\text{q}}(u, d)\), which integrates contextual and anomaly data:
\begin{equation}
    R^{\text{q}}(u, d) = \mathcal{F}_{\text{q}}(\mathbf{c}_u, \mathbf{c}_d, \mathbf{x}_i, \hat{y}_i^{\text{q}}),
\end{equation}
where \(\mathbf{c}_u\), \(\mathbf{c}_d\) are contextual features for the user and device, such as roles, permissions, and device identifiers.
$\mathbf{x}_i$ is network traffic features, e.g., packet size, connection duration, and protocol type.
 \(\hat{y}_i^{\text{q}}\) is the quantum anomaly score derived from the quantum-enhanced IDS.
 \(\mathcal{F}_{\text{q}}\) defines the quantum risk evaluation function that leverages quantum feature encoding and higher-dimensional Hilbert spaces to integrate diverse data inputs.
Access is granted if \(R^{\text{q}}(u, d)\) is below a predefined threshold and denied otherwise:
\begin{equation}
\begin{aligned}
    R^{\text{q}}(u, d) < \tau &\implies \text{Access Granted}, \\
    R^{\text{q}}(u, d) \geq \tau &\implies \text{Access Denied}.
\end{aligned}
\end{equation}
where \(\tau\) represents a predefined threshold value used for decision-making in the access control process. Specifically, \(\tau\) serves as a cutoff value for the quantum risk score \(R^{\text{q}}(u, d)\), which evaluates the potential risk of granting access to a user-device pair \((u, d)\). 
If the computed quantum risk score \(R^{\text{q}}(u, d)\) is below \(\tau\), the system considers the access request safe and grants access.
Conversely, if \(R^{\text{q}}(u, d)\) is equal to or greater than \(\tau\), access is denied or restricted due to the associated risk.
 A lower \(\tau\) increases sensitivity, making the system stricter but potentially leading to more false positives (denying legitimate access).
 A higher \(\tau\) reduces sensitivity, allowing more access but increasing the risk of admitting potential threats.
The value of \(\tau\) is optimized during the system's calibration phase by leveraging quantum-enhanced methods, such as variational quantum circuits and quantum search algorithms. Performance metrics like the True Positive Rate (\texttt{TPR}) and False Positive Rate (\texttt{FPR}) are used to determine the threshold that achieves the best balance between detecting threats and minimizing disruptions. Table~\ref{tab:math_symbols} lists the mathematical symbols used in the paper. 
\begin{figure}[h!]
    \centering
    \includegraphics[width=3.0in]{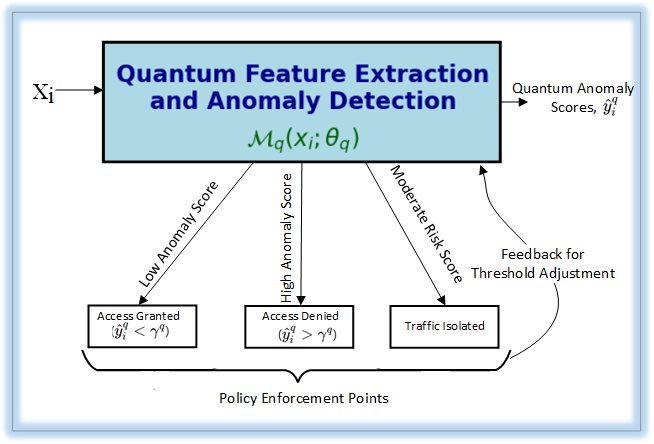}
    \caption{Quantum Feature Extraction and Anomaly Detection Layer}
    \label{fig:4}
\end{figure}

Fig.~\ref{fig:4} represents the "Quantum Feature Extraction and Anomaly Detection" layer, which processes network traffic features ($X_i$) using a quantum-enhanced model ($\mathcal{M}_q(x_i; \theta_q)$) to compute anomaly scores ($ \hat{y}_i^q $). The output anomaly scores are categorized into low, high, and moderate risk. These scores dictate actions at the Policy Enforcement Points (PEPs). Low anomaly scores result in "Access Granted," high anomaly scores trigger "Access Denied," and moderate risk scores lead to "Traffic Isolated." A feedback loop facilitates dynamic threshold adjustment to optimize system performance.
\begin{table}[h!]
\centering
\caption{Mathematical Symbols and Descriptions}
\label{tab:math_symbols}
\begin{tabular}{|p{0.85cm}|p{7.0cm}|}
\hline
\textbf{Symbol} & \textbf{Description} \\
\hline
\(f_i\) & A network flow in the system. \\
\hline
\(\mathbf{x}_i\) & Feature vector representing classical traffic attributes of  \(f_i\). \\
\hline
\(|\psi_i\rangle\) & Quantum state representation of the feature vector \(\mathbf{x}_i\). \\
\hline
\(\mathcal{M}_{\text{q}}\) & QNN model. \\
\hline
\(\hat{y}_i^{\text{q}}\) & Quantum-enhanced anomaly score computed for \(f_i\). \\
\hline
\(\theta_{\text{q}}\) & Trainable variational parameters of the QNN. \\
\hline
\(C(\theta_{\text{q}})\) & Cost function evaluated during QNN optimization. \\
\hline
\(\eta\) & Learning rate for classical optimizer. \\
\hline
\(\gamma^{\text{q}}\) & Anomaly detection threshold for classifying network flows. \\
\hline
\(R^{\text{q}}(u, d)\) & Quantum risk score computed for a user-device pair \((u, d)\). \\
\hline
\(\tau\) & Risk threshold for deciding access control in ZTF. \\
\hline
\(P(S_j)\) & Initial set of access control policies for network segment \(S_j\). \\
\hline
\(P'(S_j)\) & Dynamically adjusted access control policies for segment \(S_j\). \\
\hline
\(\mathcal{G}_{\text{q}}\) & Quantum-optimized policy adjustment function. \\
\hline
\(S_j\) & Network segment. \\
\hline
\(\texttt{FPR}_t\) & False positive rate observed at time \(t\). \\
\hline
\(\hat{\gamma}_{t+1}^{\text{q}}\) & Dynamically updated anomaly detection threshold at \(t+1\). \\
\hline
\(\Delta t\) & Grace period for delaying isolation of flagged segments. \\
\hline
\(\mathcal{D}_{FP}\) & Set of false positive samples used for QNN retraining. \\
\hline
\(\mathcal{D}_{\texttt{new}}\) & New data containing the latest attack patterns. \\
\hline
\(R^{\text{q}}(S_j)\) & Quantum-enhanced aggregated risk score for segment \(S_j\). \\
\hline
\(\text{\texttt{TPR}}\) & True Positive Rate of the anomaly detection threshold \(\gamma^{\text{q}}\). \\
\hline
\(\text{\texttt{FPR}}\) & False Positive Rate of the anomaly detection threshold \(\gamma^{\text{q}}\). \\
\hline
\(\mathcal{E}(\mathbf{x}_i)\) & Quantum encoding function. \\
\hline
\(\alpha, \beta\) & Coefficients of the quantum superposition. \\
\hline
\(|\psi_{\texttt{sup}}\rangle\) & Quantum superposition state. \\
\hline
\(|\Phi\rangle\) & Entangled quantum state correlating traffic flows. \\
\hline
\(\boldsymbol{\theta}\) & Vector of variational parameters for unitary transformation. \\
\hline
\(|\psi_{\texttt{out}}\rangle\) & Quantum-enhanced output state. \\
\hline
\(\mathcal{U}(\boldsymbol{\theta})\) & Unitary transformation parameterized by \(\boldsymbol{\theta}\). \\
\hline
\(\hat{O}\) & Observable used for calculating the quantum anomaly score. \\
\hline
\(y_q\) & Quantum anomaly score obtained by measuring \(\hat{O}\). \\
\hline
\(\mathcal{G}_{\texttt{QNN}}\) & Quantum-optimized function for updating  policies. \\
\hline
\(\gamma^{\text{q}}\) & Anomaly detection threshold used to classify network flows. \\
\hline
\(P(S_j)\) & Initial access control policy for network segment \(S_j\). \\
\hline
\(P'(S_j)\) & Dynamically updated access control policy for \(S_j\). \\
\hline
\(\hat{D}(\alpha_i)\) & Displacement operator applied to encode classical input. \\
\hline
\(\hat{S}(r_i)\) & Squeezing operator applied to enhance quantum state. \\
\hline
\(\mathcal{E}(\lambda)\) & Non-Gaussian gate introducing nonlinearity in quantum. \\
\hline
\(\phi(x)\) & Nonlinear activation function introduced by \(\mathcal{E}(\lambda)\). \\
\hline
\(M\) & Symplectic matrix representing linear transformations. \\
\hline
\(\hat{U}\) & Passive linear optical transformation in the QNN framework. \\
\hline
\(W(x, x')\) & Transformation kernel in probability amplitude mappings. \\
\hline
\(|\tilde{\psi}\rangle\) & Output wave function after a quantum transformation. \\
\hline
\(\tilde{p}(x)\) & Probability distribution from output wave function. \\
\hline
\end{tabular}
\end{table}

\subsection{Micro-segmentation}
To minimize the attack surface, ZTF incorporates micro-segmentation, which divides the network into smaller, secure segments. Each segment \(S_j\) represents a group of resources and devices with similar security requirements. By isolating these segments, ZTF ensures that lateral movement within the network is restricted even if one segment is compromised, thereby containing potential breaches.
While ZTF provides the structural framework for micro-segmentation and enforces initial access policies for each segment, the IDS plays a complementary role. The QNN-augmented IDS continuously monitors real-time network traffic within and between these segments. By leveraging quantum principles such as superposition and entanglement, the QNN can simultaneously evaluate correlations across multiple traffic flows, capturing subtle anomalies that classical systems might miss.

\textit{Quantum Anomaly Scoring in micro-segmentation:} Each flow within a segment is represented by a feature vector \(\mathbf{x}_i\), mapped into a quantum state \(|\psi_i\rangle\) using quantum feature encoding techniques such as amplitude encoding or angle encoding. Using (\ref{Eq_TVP1}), the QNN computes the quantum anomaly score \(\hat{y}_i^{\text{q}}\), which is used to evaluate the security posture of the segment:
\begin{equation}
    \hat{y}_i^{\text{q}} = \mathcal{M}_{\text{q}}(|\psi_i\rangle; \theta_{\text{q}}),
\end{equation}
where \(\mathcal{M}_{\text{q}}\) represents the variational quantum circuit, and \(\theta_{\text{q}}\) are the trainable quantum parameters.
Based on the quantum anomaly scores provided by QNN, the system dynamically adjusts the access control policies for each segment. Let \(P(S_j)\) denote the initial set of access control policies for segment \(S_j\). The IDS evaluates the traffic in real-time and uses the quantum anomaly scores \(\hat{y}_i^{\text{q}}\) to modify these policies dynamically:
\begin{equation}
    P'(S_j) = \mathcal{G}_{\text{q}}(P(S_j), \hat{y}_i^{\text{q}}),
\end{equation}
where \(\mathcal{G}_{\text{q}}\) is a quantum-optimized policy adjustment function that incorporates quantum-enhanced detection metrics.
If the QNN identifies a segment as compromised—indicated by the anomaly score exceeding a predefined quantum threshold (\(\hat{y}_i^{\text{q}} > \gamma^{\text{q}}\))—the system isolates the affected segment to prevent the threat from spreading:
\begin{equation}
    \texttt{If } \hat{y}_i^{\text{q}} > \gamma^{\text{q}}, \quad S_j \to \texttt{Isolated}.
\end{equation}

The use of QNNs in micro-segmentation provides several unique advantages, such as enabling the modeling of correlations between traffic flows across segments and allowing for early detection of coordinated attacks.
 By leveraging quantum Hilbert spaces, QNNs can more effectively represent and process high-dimensional feature spaces than classical methods.
 QNNs' quantum-enhanced computational power allows the system to rapidly adapt policies in response to evolving traffic patterns and threat landscapes.
Integrating quantum-enhanced anomaly detection and dynamic policy updates ensures that micro-segmentation effectively limits the lateral movement of attackers, containing potential breaches and maintaining the integrity of unaffected segments.
Table~\ref{Table:TauGamma} shows the difference between \(\tau\) and \(\gamma^{\text{q}}\).
\begin{table}[t]
\caption{Comparison of \(\tau\) and \(\gamma^{\text{q}}\)}
\label{Table:TauGamma}
\centering
\begin{tabular}{|p{3.5cm}|p{4.5cm}|}
\hline
~~~~~~~~~~~~~~~~\textbf{\(\tau\)} & ~~~~~~~~~~~~~~~~~\textbf{\(\gamma^{\text{q}}\)} \\
\hline
\(\tau\) is risk threshold & \(\gamma^{\text{q}}\) is anomaly detection threshold \\
\hline
Evaluates risk of  \(R^{\text{q}}(u, d)\) & Evaluates anomaly score for  \(\hat{y}_i^{\text{q}}\) \\
\hline
Access control decision & Anomaly detection in network traffic \\
\hline
Determines the access & Identifies anomalous traffic for response \\
\hline
 Applies to the global level & Applies within or across segments \\
\hline
Controls overall access policy & Controls micro-segmentation and threat \\
\hline
 Balances security and usability & Balances sensitivity and specificity  \\
\hline
\end{tabular}
\end{table}

\subsection{Threshold Optimization and Sensitivity Analysis}
In the proposed system model, the anomaly detection threshold \(\gamma^{\text{q}}\) plays a pivotal role in ensuring the effective classification of network flows \(f_i\) as either anomalous or normal. This threshold directly impacts the system's detection rate, false positive rate, and overall operational efficiency. An optimized threshold balances \textit{sensitivity} (detecting true anomalies) and \textit{specificity} (minimizing false positives). The QNN-enhanced IDS leverages quantum properties to perform efficient searches and evaluations of threshold values, ensuring robust anomaly detection under dynamic network conditions \cite{montanaro2016quantum}.

\subsection{Qubit Requirements and Scalability}
\label{sec:qubits}
QNN-based architecture proposed in this paper relies on quantum superposition, entanglement, and high-dimensional feature encoding principles. The number of qubits required in the model depends on the following factors:
Network traffic features such as packet size, connection duration, and protocol type are encoded into quantum states using amplitude or angle encoding methods. The number of qubits scales with the input features' dimensionality and the encoding resolution.
The architecture employs variational quantum circuits with trainable parameters that leverage single-qubit and two-qubit gates. The number of qubits required increases with the depth and width of these circuits, which correspond to the complexity of the anomaly detection task.
QNNs map classical data into high-dimensional Hilbert spaces, leveraging additional qubits to encode correlations and interdependencies among traffic flows.
NISQ device limitations framework is designed to operate on NISQ devices, with qubit availability typically ranging from tens to hundreds. This ensures practical scalability and compatibility with current quantum hardware.
For instance, a system encoding 50 classical features with medium precision and implementing a shallow variational circuit might require approximately 70 qubits, considering ancillary qubits for error mitigation. As the input complexity grows or deeper circuits are applied, the qubit requirement will scale accordingly.

\subsubsection{Optimization of \(\gamma^{\text{q}}\)}
The optimal threshold value, \({\gamma^{q}}^*\), is determined by maximizing quantum-enhanced performance metrics, such as the \textit{Receiver Operating Characteristic (ROC) Curve} and the \textit{Area Under the Curve (AUC)}. The ROC curve plots the \texttt{TPR} against the \texttt{FPR} for various \(\gamma^{\text{q}}\) values. The ideal threshold \({\gamma^{q}}^*\) is selected to achieve a balance between detection accuracy and operational efficiency:
\begin{equation}
    {\gamma^{q}}^* = \arg\max_{\gamma^{\text{q}}} \left( \text{\texttt{TPR}}(\gamma^{\text{q}}) - \text{\texttt{FPR}}(\gamma^{\text{q}}) \right),
\end{equation}
where
\begin{align}
    \text{\texttt{TPR}}(\gamma^{\text{q}}) &= \frac{\texttt{True Positives}}{\texttt{True Positives} + \texttt{False Negatives}}, \nonumber \\
    \text{\texttt{FPR}}(\gamma^{\text{q}}) &= \frac{\texttt{False Positives}}{\texttt{False Positives} + \texttt{True Negatives}}. \nonumber
\end{align}
The threshold \(\gamma^{\text{q}}\) is fine-tuned using cross-validation on labeled datasets containing normal and anomalous traffic patterns. Quantum-enhanced search algorithms, such as Grover’s search, accelerate the optimization process, providing faster convergence compared to classical methods \cite{montanaro2016quantum}. This ensures the QNN-augmented IDS performs consistently and accurately across diverse traffic scenarios, including emerging attack patterns.

\subsubsection{Sensitivity Analysis}
Sensitivity analysis evaluates the robustness of the system by analyzing how variations in \(\gamma^{\text{q}}\) affect performance. The system examines the detection rates and false positive rates over a range of \(\gamma^{\text{q}}\) values to identify the threshold that provides the best trade-off. A small change in the threshold, denoted as \(\Delta \gamma^{\text{q}}\), affects system performance as follows:
\begin{equation}
    \text{Sensitivity} = \frac{\partial \text{\texttt{TPR}}}{\partial \gamma^{\text{q}}} - \frac{\partial \text{\texttt{FPR}}}{\partial \gamma^{\text{q}}}.
\end{equation}

The system dynamically adapts \(\gamma^{\text{q}}\) in real time to accommodate evolving traffic patterns and emerging threats using quantum-enhanced optimization methods. This dynamic adaptability ensures that the QNN-augmented IDS effectively detects anomalies while minimizing disruptions to legitimate traffic \cite{cerezo2022challenges}.

\subsubsection{Qubit Estimation for Feature Encoding}
Let $N_f$ represent the number of classical features encoding each network flow. For quantum feature encoding methods such as amplitude or angle encoding, the number of qubits required is given by:
\begin{equation}
Q_{\texttt{enc}} = \lceil \log_2(N_f) \rceil,
\end{equation}
where $Q_{\texttt{enc}}$ denotes the number of qubits needed to represent the feature space, and $\lceil \cdot \rceil$ is the ceiling function.

\subsubsection{Qubit Estimation for Variational Quantum Circuits}
Let $L$ denote the number of variational layers in the circuit, and $Q_{\texttt{gate}}$ the qubits required for entanglement and gate operations per layer. The total qubits required for variational circuits are:
\begin{equation}
Q_{\texttt{vqc}} = Q_{\texttt{enc}} + L \cdot Q_{\texttt{gate}},
\end{equation}
where $Q_{\texttt{gate}}$ depends on the type of gates (e.g., \texttt{CNOT}, \texttt{SWAP}) and the circuit topology.

\subsubsection{Overall Qubit Estimation}
The total number of qubits, $Q_{\texttt{total}}$, is estimated as:
\begin{equation}
Q_{\texttt{total}} = Q_{\texttt{enc}} + Q_{\texttt{vqc}},
\end{equation}
where $Q_{\texttt{enc}}$ accounts for feature encoding, and $Q_{\texttt{vqc}}$ incorporates the qubits needed for variational circuit operations.
For practical deployment on NISQ devices, the number of qubits is constrained by device capabilities, typically 50 to 100 qubits. This scalability ensures efficient implementation of the QNN-ZTF across diverse network environments.

\subsubsection{Scalability and Adaptation}
The proposed framework dynamically adjusts $N_f$ to optimize qubit utilization by selecting the most critical traffic features using feature selection techniques. Additionally, the number of variational layers $L$ can be adapted based on the desired trade-off between accuracy and computational resources.

\subsection{Handling of False Positives and Negatives}
Unlike classical systems, the proposed model leverages QNN retraining with quantum feature encoding, including techniques such as amplitude, angle, and basis encoding, to improve anomaly classification accuracy. The model exploits quantum superposition and interference by encoding classical traffic attributes (e.g., packet size, connection duration, source/destination IPs, and protocol type) into quantum states to enhance the representation of complex patterns. This approach enables the QNN to recalibrate its variational parameters \(\theta_{\text{q}}\) more efficiently, reducing false positive and negative rates over time. The iterative retraining process dynamically adapts the anomaly detection system to evolving network conditions, ensuring robustness and operational efficiency.
 A feedback loop collects misclassified samples, which are re-encoded into quantum states and used to refine the variational parameters \(\theta_{\text{q}}\). This process exploits quantum superposition and interference to recalibrate the model efficiently, reducing false positive and negative rates over time.

\subsubsection{Minimizing False Positives}
False positives occur when legitimate traffic is misclassified as anomalous, leading to unnecessary isolation of network segments and disruptions to legitimate operations. The system mitigates false positives using the following multi-layered approach:

\textit{Feedback Loop for QNN Retraining}: 
The proposed model employs a feedback loop that leverages quantum feature encoding to improve anomaly classification accuracy. False positive samples are collected and represented as:
\begin{equation}
\mathcal{D}_{FP} = \{f_i \mid \hat{y}_i^{\text{q}} > \gamma^{\text{q}}, y_i = 0\},
\end{equation}
where \(\hat{y}_i^{\text{q}}\) is the predicted anomaly score generated by the QNN.
\(\gamma^{\text{q}}\) is the anomaly detection threshold.
 \(y_i = 0\) is the legitimate traffic incorrectly classified as anomalous.
These false positive samples are re-encoded into quantum states using quantum feature encoding techniques, such as amplitude or angle encoding. They are used to retrain the QNN anomaly detection model \(\mathcal{M}_{\text{q}}\). The retraining process refines the variational parameters \(\theta_{\text{q}}\), enhancing the model's ability to differentiate between normal and anomalous traffic:
\begin{equation}
\label{Eq_QSI}
\mathcal{M}_{\text{q}} \leftarrow \mathcal{M}_{\text{q}} + \mathcal{D}_{FP}.
\end{equation}

Using (\ref{Eq_QSI}), the model can efficiently adapt to evolving network conditions, minimizing false positive rates while maintaining high detection accuracy. This iterative process ensures the IDS remains robust and effective over time.

\textit{Dynamic Threshold Adjustment}: 
In the proposed model, the anomaly detection threshold \(\gamma^{\text{q}}\) is dynamically updated using quantum-enhanced feedback mechanisms. This ensures the system adapts to evolving network conditions and traffic patterns while maintaining high detection accuracy. Let \(\hat{\gamma}_{t+1}^{\text{q}}\) represent the threshold at time \(t+1\). The updated threshold minimizes the false positive rate (\(\texttt{FPR}\)) using a quantum-optimized feedback function \(\mathcal{F}_{\gamma}^{\text{q}}\):
\begin{equation}
\hat{\gamma}_{t+1}^{\text{q}} = \mathcal{F}_{\gamma}^{\text{q}}(\hat{\gamma}_t^{\text{q}}, \texttt{FPR}_t),
\end{equation}
where \(\hat{\gamma}_t^{\text{q}}\) is the threshold value at the current time step \(t\).
 \(\texttt{FPR}_t\) is the false positive rate observed at time \(t\), calculated from quantum-enhanced anomaly scores.
\(\mathcal{F}_{\gamma}^{\text{q}}\) is a quantum-augmented feedback function that leverages quantum feature encoding and interference to optimize threshold adjustments.
By utilizing quantum superposition and interference, the feedback function \(\mathcal{F}_{\gamma}^{\text{q}}\) evaluates multiple potential adjustments simultaneously, accelerating the convergence to an optimal threshold.

\textit{Grace Periods}: 
For segments flagged as abnormal, isolation is delayed by a grace period \(\Delta t\) to allow for additional verification checks:
\begin{equation}
\text{Isolation}(S_j) \text{ if } t > \Delta t \text{ and } \hat{y}_i^{\text{q}} > \gamma^{\text{q}}.
\end{equation}

\subsubsection{Managing False Negatives}
False negatives occur when anomalous traffic is misclassified as normal, potentially allowing malicious activity undetected. The system mitigates false negatives with the following strategies:

\textit{Aggressive Detection in High-Risk Scenarios}: 
For segments with high aggregated risk scores \(R^{\text{q}}(S_j)\), the threshold \(\gamma^{\text{q}}\) is lowered to increase sensitivity:
\begin{equation}
\gamma^{\text{q}} \rightarrow \gamma^{\prime \text{q}} \quad \texttt{If } R^{\text{q}}(S_j) > \tau,
\end{equation}
where \(\gamma^{\text{q}}\) is current anomaly detection threshold.
  \(\gamma^{\prime \text{q}}\) is adjusted threshold to increase sensitivity.
 \(R^{\text{q}}(S_j)\) is quantum-enhanced risk score for segment \(S_j\), derived from contextual data and anomaly scores.
  \(\tau\) is a predefined risk threshold identifying high-risk scenarios.
When \(R^{\text{q}}(S_j)\) exceeds \(\tau\), the system dynamically lowers \(\gamma^{\text{q}}\) to \(\gamma^{\prime \text{q}}\), enhancing its sensitivity to potential threats.

\textit{Continuous QNN Updates:} 
The anomaly detection model \(\mathcal{M}_{\text{q}}\) is periodically updated using new data \(\mathcal{D}_{\texttt{new}}\) containing the latest attack patterns:
\begin{equation}
\label{Eq_MQNN}
\mathcal{M}_{\text{q}} \leftarrow \mathcal{M}_{\text{q}} + \mathcal{D}_{\texttt{new}}.
\end{equation}

In (\ref{Eq_MQNN}), the model remains robust against emerging threats and continues to refine its anomaly classification capabilities.
\begin{figure*}[h!]
    \centering
    \includegraphics[width=6.0in]{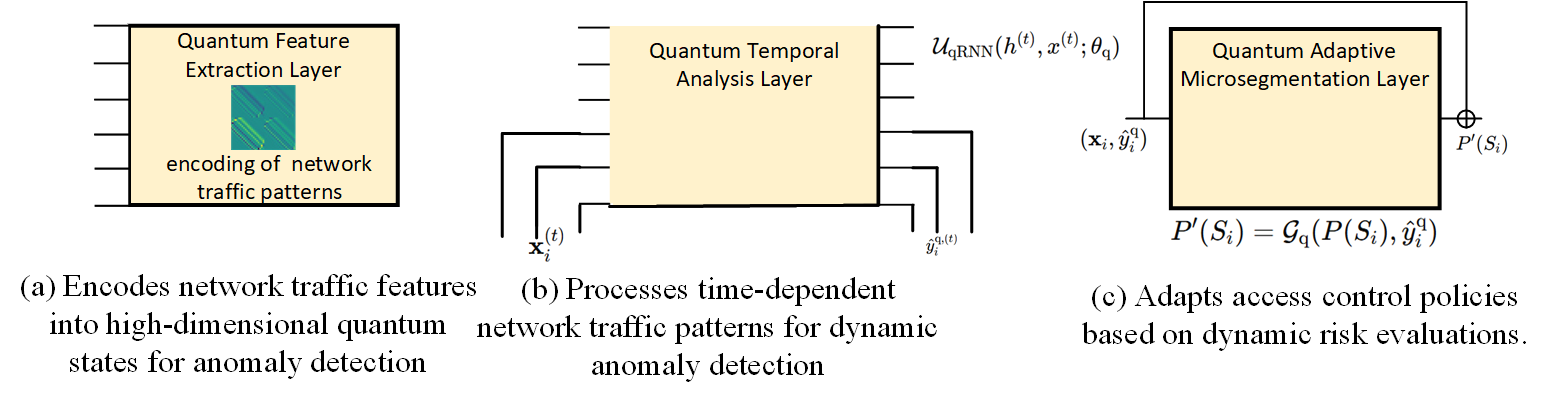}
    \caption{Architecture of the Quantum-Enhanced ZTF}
    \label{fig:3}
\end{figure*}

Fig.~\ref{fig:3} illustrates the key components of the quantum-enhanced anomaly detection framework. In Fig.~\ref{fig:3}(a), the Quantum Feature Extraction Layer encodes network traffic features into high-dimensional quantum states, enabling enhanced anomaly detection by leveraging quantum superposition and entanglement. Fig.~\ref{fig:3}(b) depicts the Quantum Temporal Analysis Layer, which processes time-dependent network traffic patterns using quantum recurrent analysis, allowing for dynamic detection of evolving threats. Finally, Fig.~\ref{fig:3}(c) showcases the Quantum Adaptive Micro-Segmentation Layer, which dynamically adjusts access control policies based on quantum-enhanced risk evaluations, ensuring real-time containment of threats and limiting lateral movement within the network. This layered quantum approach significantly improves the responsiveness and accuracy of anomaly detection in cybersecurity environments.

\subsection{Simultaneous Operation of ZTF and IDS}
ZTF and IDS operate simultaneously in the proposed system, enabling dynamic and adaptive threat response. ZTF computes the risk score \(R^{\text{q}}(u, d)\) based on contextual and traffic data and quantum anomaly scores from IDS \cite{cerezo2022challenges}. If \(R^{\text{q}}(u, d) < \tau\), access is granted. Otherwise, access is denied or restricted. Simultaneously, QNN-enhanced IDS continuously monitors the behavior of granted entities, updating anomaly scores \(\hat{y}_i^{\text{q}}\) in real-time. These scores inform ZTF to dynamically adjust access policies, isolate suspicious entities, or revoke access as needed. This simultaneous operation ensures robust security by combining the quantum-enhanced monitoring capabilities of IDS with the dynamic enforcement of ZTF policies. By leveraging preventive, detective, and quantum-enhanced measures, the system achieves a comprehensive security posture \cite{rose2019zero}.
The proposed QNN framework is designed to scale effectively with large, complex networks. By leveraging quantum-enhanced feature encoding, the system can process high-dimensional input spaces without incurring significant computational overhead. The quantum-classical hybrid approach allows efficient parameter optimization even with increasing network size. This scalability ensures the system remains robust and responsive in environments with large-scale traffic flows, diverse device types, and rapidly evolving threat landscapes.

\section{Proposed QNN-based Solution} \label{Solution}
This section integrates the QNN paradigm into ZTF with dynamic anomaly detection and adaptive micro-segmentation. The proposed approach leverages QNNs to enhance the detection of anomalies and the dynamic enforcement of micro-segmentation policies. By incorporating the quantum principles of superposition, entanglement, and quantum feature encoding, QNNs enable higher-dimensional representations, leading to precise and adaptive security measures.
This framework takes inspiration from classical neural networks, which act as universal function approximators and have demonstrated robust performance across diverse problem domains. Additionally, it draws from the structure of variational quantum circuits, which represent a robust methodology for quantum optimization and learning on near-term quantum devices \cite{killoran2019continuous, cerezo2022challenges}. Table~\ref{tab:ztf_cv_correspondences} provides a detailed mapping between classical neural network concepts and their QNN equivalents, contextualized for the ZTF.
\begin{figure*}[h!]
    \centering
    \includegraphics[width=7.0in]{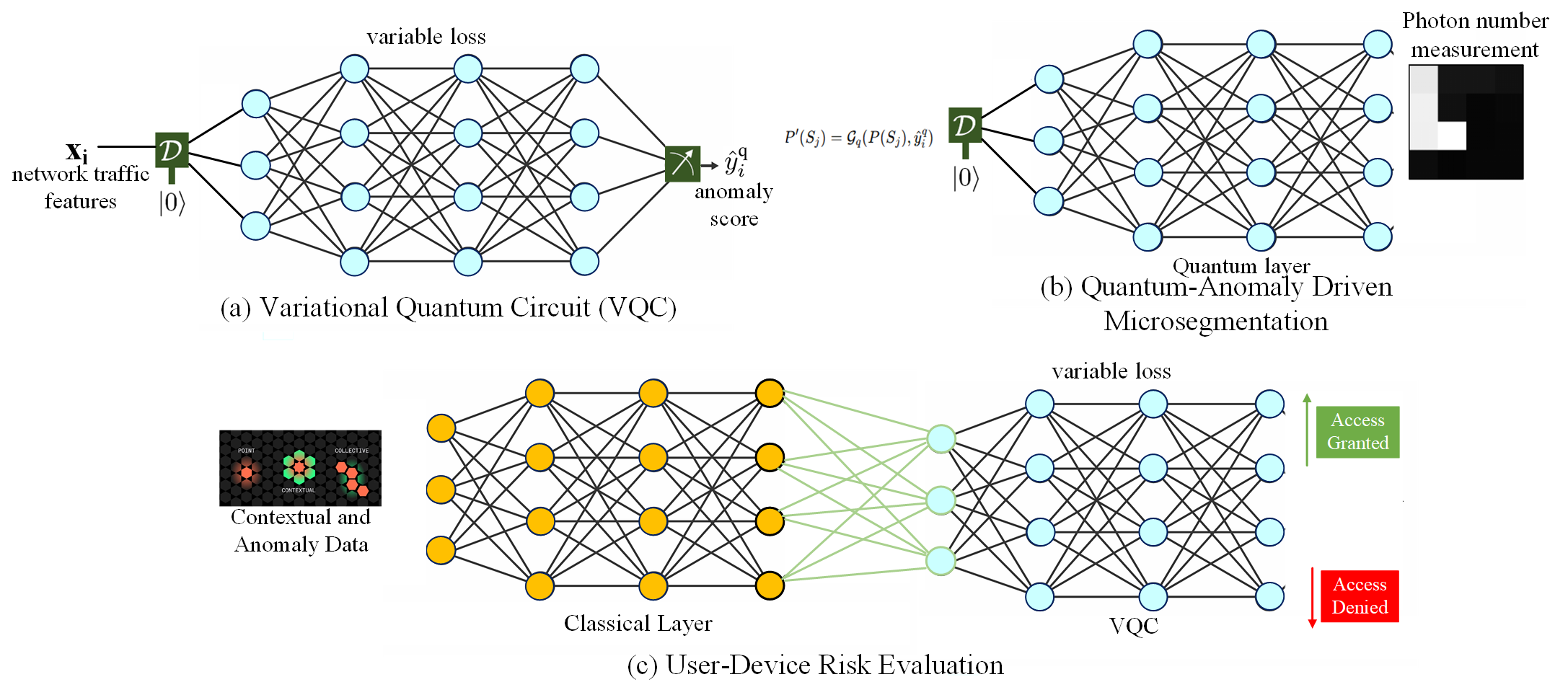}
\caption{Illustration of the Quantum-Enhanced ZTF with dynamic anomaly detection, adaptive micro-segmentation, and risk-based access control.}
\label{fig:quantum_zero_trust}
\end{figure*}

Fig.~\ref{fig:quantum_zero_trust} illustrates the integration of the Quantum-Enhanced ZTF for dynamic anomaly detection and adaptive micro-segmentation using VQC. In Fig.~\ref{fig:quantum_zero_trust}(a), network traffic features ($x_i$) are encoded into quantum states using displacement gates ($D$), and processed through a variational quantum circuit to compute quantum-enhanced anomaly scores ($\hat{y}_i^q$). These scores dynamically guide segmentation policy adjustments, represented as $P'(S_j) = \mathcal{G}_q(P(S_j), \hat{y}_i^q)$. Fig.~\ref{fig:quantum_zero_trust}(b) demonstrates quantum-anomaly-driven micro-segmentation, where quantum layers analyze high-dimensional traffic patterns and compute anomaly metrics, with photon number measurements supporting segmentation adjustments. Fig.~\ref{fig:quantum_zero_trust}(c) highlights user-device risk evaluation, where contextual and anomaly data are processed through classical and quantum layers to compute quantum risk scores ($R^q(u, d)$). Access decisions are then based on the computed scores, granting access to low-risk evaluations and denying access to high-risk evaluations. The framework effectively combines quantum principles, such as superposition and entanglement, to enhance anomaly detection, enforce micro-segmentation, and enable dynamic, risk-based access control.
\begin{table}[ht]
    \centering
    \caption{Correspondences between classical NN and QNN}
    \label{tab:ztf_cv_correspondences}
    \begin{tabular}{|p{2.47cm}|p{5.550cm}|}
        \hline
        \multicolumn{1}{|c|}{\textbf{Classical NN}} & \multicolumn{1}{c|}{\textbf{QNN}} \\
        \hline
        Weight matrix & Symplectic matrix (security encoding) \\ \hline
        Bias vector \(b\) & Displacement vector \(\alpha\) (policy adjustment) \\ \hline
        Affine transformation & Gaussian gates (risk evaluation) \\ \hline
        Nonlinear function & Non-Gaussian gates (anomaly scoring) \\ \hline
        Weight/bias & Gate parameters optimized via variational circuits \\ \hline
        Variable \(x\) & Operator \(\hat{x}\) (traffic attributes) \\ \hline
        Derivative \(\frac{\partial}{\partial x}\) & Conjugate operator \(\hat{p}\) (response gradient) \\ \hline
        \textit{No classical analog} & Superposition (simultaneous traffic evaluation) \\ \hline
        \textit{No classical analog} & Entanglement (correlated anomaly patterns) \\ 
        \hline
    \end{tabular}
\end{table}

\subsubsection{Quantum Feature Encoding}
Classical input features \(\mathbf{x}_i \in \mathbb{R}^n\), representing traffic attributes such as source/destination IPs, packet size, connection duration, and protocol type, are mapped into a quantum state \(|\psi_i\rangle\) in a higher-dimensional Hilbert space \(\mathcal{H}\). This mapping is achieved through a quantum encoding function \(\mathcal{E}\), enabling the system to capture richer correlations among features that are challenging for classical methods to detect:
\begin{equation}
|\psi_i\rangle = \mathcal{E}(\mathbf{x}_i),
\end{equation}
where \(\mathcal{E}\) is the quantum encoding function, such as amplitude or angle encoding, capturing richer correlations between input features.

\subsubsection{Quantum Superposition}
Quantum superposition enables the simultaneous representation and processing of multiple input states, significantly enhancing computational parallelism. For two feature vectors, \(\mathbf{x}_1\) and \(\mathbf{x}_2\), their superposition in the Hilbert space \(\mathcal{H}\) is represented as:
\begin{equation}
|\psi_{\texttt{sup}}\rangle = \alpha |\psi_1\rangle + \beta |\psi_2\rangle,
\end{equation}
where \(|\psi_1\rangle = \mathcal{E}(\mathbf{x}_1)\) and \(|\psi_2\rangle = \mathcal{E}(\mathbf{x}_2)\) are the quantum states corresponding to the encoded feature vectors, and \(\alpha, \beta \in \mathbb{C}\) are coefficients satisfying the normalization condition\(|\alpha|^2 + |\beta|^2 = 1\), which allows parallel evaluation of security metrics for different data flows.

\subsubsection{Entanglement}
Entanglement introduces correlations between subsystems in a multipartite quantum state. For example, if two qubits encode traffic flows \(f_1\) and \(f_2\), the entangled state is:
\begin{equation}
|\Phi\rangle = \frac{1}{\sqrt{2}} \left(|\psi_{f_1}\rangle \otimes |\psi_{f_2}\rangle + |\psi_{f_2}\rangle \otimes |\psi_{f_1}\rangle \right).
\end{equation}
Such correlations enhance anomaly detection by uncovering relationships between features across different traffic flows.

\subsubsection{Quantum Variational Optimization}
A variational circuit parameterized by \(\boldsymbol{\theta}\) acts on the encoded quantum state to generate a quantum-enhanced representation:
\begin{equation}
|\psi_{\texttt{out}}\rangle = \mathcal{U}(\boldsymbol{\theta})|\psi_i\rangle,
\end{equation}
where \(\mathcal{U}(\boldsymbol{\theta})\) is a unitary transformation defined by the quantum circuit. The output state \(|\psi_{\texttt{out}}\rangle\) encodes higher-dimensional relationships that are optimized during training.

\subsubsection{Adaptive Security Decisions}
The quantum anomaly score \(y_q\) is calculated by measuring an observable \(\hat{O}\) on the output state:
\begin{equation}
y_q = \langle \psi_{\texttt{out}} | \hat{O} | \psi_{\texttt{out}} \rangle,
\end{equation}
where \(\hat{O}\) is a problem-specific operator. Based on the score \(y_q\), access control policies are dynamically updated:
\begin{equation}
P'(S_j) = \mathcal{G}_{\texttt{QNN}}(P(S_j), y_q),
\end{equation}
where \(P'(S_j)\) represents the updated policy for segment \(S_j\), and \(\mathcal{G}_{\texttt{QNN}}\) is a quantum-optimized function incorporating anomaly detection metrics.
In the proposed ZTF, QNNs are employed to compute quantum-enhanced anomaly scores for network traffic. Each traffic flow \(f_i\) is represented by a feature vector \(\mathbf{x}_i\), encoding attributes such as source/destination IPs, packet sizes, connection durations, and protocol types. The QNN transforms these features into quantum states using quantum feature encoding techniques such as amplitude or angle encoding. This quantum representation enables the detection of complex correlations and dependencies that are challenging to identify using classical methods.
The anomaly score \(\hat{y}_i^{\text{q}}\) is calculated using a variational quantum circuit optimized for high-dimensional traffic patterns:
\begin{equation}
    \hat{y}_i^{\text{q}} = \mathcal{M}_{\text{q}}(\mathbf{x}_i; \theta_{\text{q}}),
\end{equation}
where \(\theta_{\text{q}}\) is the variational parameters trained through a hybrid quantum-classical optimization loop. The threshold \(\gamma^{\text{q}}\) is dynamically adjusted to maintain a balance between sensitivity and specificity, ensuring accurate anomaly detection.
To enhance micro-segmentation, the ZTF leverages quantum anomaly scores to dynamically update access policies within and between network segments. Each segment \(S_j\) is monitored using QNN-based anomaly detection. The QNN computes a segment-specific anomaly score, \(\hat{y}_i^{\text{q}}\), which informs policy adjustments:
\begin{equation}
    P'(S_j) = \mathcal{G}_{\text{q}}(P(S_j), \hat{y}_i^{\text{q}}),
\end{equation}
where \(\mathcal{G}_{\text{q}}\) represents a quantum-optimized function for policy updates. Segments flagged with anomalies exceeding the threshold (\(\hat{y}_i^{\text{q}} > \gamma^{\text{q}}\)) are isolated to prevent lateral movement of potential threats:
\begin{equation}
    \texttt{If } \hat{y}_i^{\text{q}} > \gamma^{\text{q}}, \quad S_j \to \texttt{Isolated}.
\end{equation}

Integrating QNNs into ZTF offers significant advantages by leveraging quantum superposition to evaluate multiple traffic flows simultaneously, enabling the detection of subtle and complex abnormal behavior patterns. Quantum-enhanced feedback loops facilitate real-time updates to segmentation policies, dynamically adapting to evolving traffic patterns and improving threat detection. By isolating high-risk segments, the system effectively prevents the lateral propagation of threats while maintaining operational continuity. The quantum-classical hybrid approach ensures efficient handling of the computational demands associated with large-scale networks, making the framework scalable and practical. This integration establishes a robust, adaptive, quantum-optimized defense mechanism that addresses modern network security challenges.

\subsection{Fully Connected Quantum Layers}
The ZTF integrates QNNs as a sequence of fully connected quantum layers. Each layer consists of a structured sequence of operations designed to transform network traffic data into high-dimensional quantum representations for dynamic anomaly detection and policy enforcement. These layers enable the system to efficiently model and respond to complex network patterns and security threats.
A general layer, \(L\), is defined as:
\begin{equation}
    L := \mathcal{E} \circ D \circ U_2 \circ S \circ U_1,
\end{equation}
where  \(U_i = U_i(\theta_i, \phi_i)\)is general \(N\)-port linear optical interferometers consisting of beamsplitter and rotation gates.
\(D = \bigotimes_{i=1}^{N} \hat{D}(\alpha_i)\) is collective displacement operators applied to each mode.
\(S = \bigotimes_{i=1}^{N} \hat{S}(r_i)\) is collective squeezing operators applied independently to each mode.
 \(\mathcal{E} = \mathcal{E}(\lambda)\) is a non-Gaussian gate, such as a cubic phase gate or Kerr gate, providing nonlinearity and universality.
The parameters \((\theta, \phi, r, \boldsymbol{\alpha}, \lambda)\) are the trainable variables of the QNN and represent quantum analogs of weights and biases in classical neural networks. These parameters are optimized to minimize errors in detecting anomalous traffic patterns and enforcing micro-segmentation policies.
In the ZTF, fully connected quantum layers are employed to compute anomaly scores for network flows. The transformation applied by a single layer \(L\) can be expressed as:
\begin{equation}
    L(\hat{\mathbf{r}}) = \phi(M\hat{\mathbf{r}} + \boldsymbol{\alpha}),
\end{equation}
where \(\hat{\mathbf{r}} = (\hat{x}, \hat{p})\)is a vector of canonical quantum operators representing encoded traffic attributes (e.g., packet size, connection duration).
 \(M\) is a symplectic matrix encoding linear transformations analogous to classical weight matrices.
  \(\boldsymbol{\alpha}\): A displacement vector analogous to biases, enabling adaptive policy updates.
  \(\phi\) is a nonlinear function introduced by the non-Gaussian gate \(\mathcal{E}\), providing universality for anomaly detection tasks.
This quantum transformation encodes traffic flows into high-dimensional quantum feature spaces, enabling the detection of complex correlations between traffic patterns. For example, using the cubic-phase gate, the nonlinearity \(\phi\) can be represented as:
\begin{equation}
    \phi(V(\lambda)) :
    \begin{bmatrix}
        \hat{x} \\ 
        \hat{p}
    \end{bmatrix}
    \to
    \begin{bmatrix}
        \hat{x} \\ 
        \hat{p} + \lambda \hat{x}^2
    \end{bmatrix},
\end{equation}
where \(\lambda\) is a tunable parameter of the cubic-phase gate.
Stacking multiple fully connected quantum layers enables deeper modeling of network behavior, facilitating adaptive micro-segmentation. The output quantum state of one layer serves as the input to the next, with additional qumodes added or removed between layers to adjust the network's width. This architecture supports dynamic segmentation by encoding classical network attributes (e.g., source/destination IPs) as quantum states through displacement operations \(D(\boldsymbol{\alpha})\), applying symplectic transformations \(U_1, U_2\) to capture linear correlations between flows, and leveraging non-Gaussian gates \(\mathcal{E}\) to detect non-linear relationships indicative of anomalous behavior or coordinated attacks.
For example, if the QNN identifies anomalous behavior in a segment \(S_j\), the policy \(P(S_j)\) is dynamically updated using quantum-enhanced insights:
\begin{equation}
    P'(S_j) = \mathcal{G}_{\text{q}}(P(S_j), \hat{y}_i^{\text{q}}),
\end{equation}
where \(\mathcal{G}_{\text{q}}\) is a quantum-optimized policy adjustment function informed by anomaly scores \(\hat{y}_i^{\text{q}}\).
The QNN framework supports hybrid input schemes, allowing classical and quantum data to be processed seamlessly. Classical inputs, such as contextual access control features, are embedded into the quantum feature space using displacement operations:
\begin{equation}
    D(\mathbf{x}_i)|0\rangle,
\end{equation}
where \(\mathbf{x}_i\) represents the classical input feature vector. This hybrid processing enables the framework to combine quantum anomaly detection capabilities with classical contextual information for more accurate decision-making.
The output of each fully connected quantum layer can be analyzed through flexible measurement techniques, including homodyne, heterodyne, or photon-counting methods. These measurements yield expectation values to compute a cost function for optimizing the QNN. Minimizing this cost function ensures the accurate detection of anomalies and optimal enforcement of micro-segmentation policies.
Using fully connected quantum layers provides several key advantages, such as non-Gaussian gates enabling universal function approximation, which is critical for detecting complex attack patterns.
Layered architectures adapt dynamically to varying network sizes and segmentation requirements.
Classical data embedding ensures seamless integration with traditional security systems.
High-dimensional quantum feature encoding enhances the sensitivity of anomaly detection.

\subsection{Embedding Classical NN into Quantum}
The QNN framework proposed in the ZTF is sufficiently general to encompass classical neural networks as a special case. This embedding is a mathematical foundation for integrating classical architectures within the quantum formalism, providing a seamless transition between classical and quantum-enhanced anomaly detection mechanisms. Notably, the embedding avoids creating superposition or entanglement in the position basis, enabling compatibility with classical computation principles.

\subsubsection{Encoding Classical Network Traffic Data}
Classical network traffic attributes (e.g., source/destination IPs, packet sizes) are represented in the quantum framework as \(N\)-mode quantum optical states, derived from eigenstates \(|x_i\rangle\) of the operators \(\hat{x}_i\):
\begin{equation}
    \mathbf{x} \leftrightarrow |\mathbf{x}\rangle := |x_1\rangle \otimes \cdots \otimes |x_N\rangle.
\end{equation}
For the first layer in a network, classical input \(\mathbf{x}\) is prepared by applying a displacement operator \(D(\mathbf{x})\) to the vacuum state \(|\mathbf{x} = 0\rangle\):
\begin{equation}
    |\mathbf{x}\rangle := D(\mathbf{x})|0\rangle.
\end{equation}
Subsequent layers use the previous layer's output as input, and the final output is measured using homodyne detection to project onto the states \(|\mathbf{x}\rangle\). This process enables the transformation of classical network traffic data into quantum states for further processing.

\subsubsection{Affine Transformations with Gaussian Operations}
The quantum embedding allows for affine transformations of classical inputs, equivalent to the feedforward operations in classical neural networks:
\begin{equation}
    |\mathbf{x}\rangle \to |W \mathbf{x} + \mathbf{b}\rangle,
\end{equation}
where \(W\) is a weight matrix, and \(\mathbf{b}\) is a bias vector. This is achieved through a sequence of Gaussian operations that include:

\textit{Orthogonal Multiplication:} The unitary operator \(U_1\) acts as a block-diagonal interferometer implementing an orthogonal matrix \(O_1\):
    \begin{equation}
        U_1|\mathbf{x}\rangle = |O_1 \mathbf{x}\rangle.
    \end{equation}

\textit{Scaling by Singular Values:} Squeezing operators \(S(r)\) scale the eigenstates by their singular values:
    \begin{equation}
        S(r)|x_i\rangle = |c_i x_i\rangle, \quad c_i = e^{-r_i}.
    \end{equation}
    Collectively, these operations apply a diagonal scaling matrix \(\Sigma\), corresponding to the singular values of \(W\).

\textit{Bias Addition:} Displacement operators \(D(\boldsymbol{\alpha})\) add the bias vector \(\mathbf{b}\) by translating the position quadrature:
    \begin{equation}
        D(\boldsymbol{\alpha})|\mathbf{x}\rangle = |\mathbf{x} + \sqrt{2}\boldsymbol{\alpha}\rangle.
    \end{equation}

Combining these operations, the sequence \(D \circ U_2 \circ S \circ U_1\) performs the desired affine transformation:
\begin{equation}
    D \circ U_2 \circ S \circ U_1|\mathbf{x}\rangle = |W \mathbf{x} + \mathbf{b}\rangle,
\end{equation}
where \(W = O_2 \Sigma O_1\) is derived from the singular value decomposition of the weight matrix.

\subsubsection{Nonlinear Transformations with Non-Gaussian Operations}
To introduce nonlinearity—an essential component for anomaly detection—the framework incorporates a non-Gaussian gate \(\mathcal{E}\), which acts element-wise on each mode:
\begin{equation}
    \mathcal{E}|\mathbf{x}\rangle = |\phi(\mathbf{x})\rangle,
\end{equation}
where \(\phi(\mathbf{x})\) is a nonlinear activation function (e.g., polynomial or piecewise-defined). This is implemented using ancillary modes and unitary operations as follows:
\begin{equation}
    |\mathbf{x}\rangle \rightarrow |\mathbf{x}\rangle \otimes |0\rangle_{\texttt{anc}},
\end{equation}

Now we re-write followed by a transformation:
\begin{equation}
    V_\phi = \exp\left(i \phi(\hat{\mathbf{x}}) \otimes \hat{p}_{\texttt{anc}}\right).
\end{equation}

The ancilla is traced out or measured, leaving the transformed state:
\begin{equation}
    |\mathbf{x}\rangle \to |\phi(\mathbf{x})\rangle.
\end{equation}

This non-Gaussian transformation enables the quantum network to approximate any function with a convergent Taylor series, extending the representational capacity beyond that of classical networks.

\subsection{Linear Interferometers}
This section derives the transformation effect of a passive linear interferometer on the eigenstates \(|x\rangle\). The foundation for this derivation begins with the expression for an eigenstate of the \(\hat{x}\) quadrature with eigenvalue \(x\):
\begin{equation}
    |x\rangle = \pi^{-1/4} \exp\left(-\frac{1}{2}x^2 + \sqrt{2}x \hat{a}^\dagger - \frac{1}{2}(\hat{a}^\dagger)^2\right)|0\rangle,
\end{equation}
where \(\hat{a} = \frac{1}{\sqrt{2}}(\hat{x} + i\hat{p})\) is the bosonic annihilation operator, and \(|0\rangle\) represents the single-mode vacuum state. Importantly, the derivation is independent of any prefactor conventions used to define the quadrature operator \(\hat{x}\) in terms of \(\hat{a}\) and \(\hat{a}^\dagger\).

\subsubsection{Multimode Generalization}
For \(N\)-mode systems, the expression generalizes as:
\begin{align}
    &|x\rangle  = \bigotimes_{i=1}^N |x_i\rangle \nonumber \\
    & = \pi^{-N/4} \exp\left(-\frac{1}{2} x^T x + \sqrt{2} x^T \hat{a}^\dagger - \frac{1}{2} (\hat{a}^\dagger)^T \hat{a}^\dagger \right) |0\rangle,
\end{align}
where \(x = (x_1, \ldots, x_N)^T\), 
\(\hat{a}^\dagger = (\hat{a}_1^\dagger, \ldots, \hat{a}_N^\dagger)^T\), and 
\(|0\rangle\) is the multimode vacuum state.

\subsubsection{Linear Optical Transformations}
Under a passive linear optical transformation \(\hat{U}\), the creation operator transforms as:
\begin{equation}
    \hat{U} \hat{a}_i^\dagger \hat{U}^\dagger = \sum_j U_{ij} \hat{a}_j^\dagger,
\end{equation}
or in matrix form:
\begin{equation}
    \hat{a}^\dagger \to U \hat{a}^\dagger, \quad (\hat{a}^\dagger)^T \to (\hat{a}^\dagger)^T U^T,
\end{equation}
where \(U\) is a unitary matrix satisfying \(U U^\dagger = I_N\).

\subsubsection{Effect on Multimode States}
We now examine the transformation of the multimode state \(|x\rangle\) under the linear optical transformation \(\hat{U}\):
\begin{align}
    \hat{U}|x\rangle = &\hat{U} \pi^{-N/4} \exp\left(-\frac{1}{2} x^T x + \sqrt{2} x^T \hat{a}^\dagger - \frac{1}{2} (\hat{a}^\dagger)^T \hat{a}^\dagger \right)|0\rangle \nonumber \\
    = & \pi^{-N/4} \exp\bigg(-\frac{1}{2} x^T x + \sqrt{2} x^T (U \hat{a}^\dagger) \nonumber \\ 
    & - \frac{1}{2} (U \hat{a}^\dagger)^T (U \hat{a}^\dagger)\bigg) \hat{U}|0\rangle.
\end{align}

Using the unitarity property \(U^T U = I_N\), the expression simplifies to:
\begin{align}
    \hat{U}|x\rangle & = \pi^{-N/4} \exp\left(-\frac{1}{2} x^T x + \sqrt{2} x^T U \hat{a}^\dagger - \frac{1}{2} (\hat{a}^\dagger)^T \hat{a}^\dagger \right)|0\rangle.
\end{align}

\subsubsection{Transformation to New Basis}
Defining \(y = U^T x\) and the orthogonal matrix \(C = U^T\), the above expression becomes:
\begin{align}
    \hat{U}|x\rangle & = \pi^{-N/4} \exp\left(-\frac{1}{2} y^T y + \sqrt{2} y^T \hat{a}^\dagger - \frac{1}{2} (\hat{a}^\dagger)^T \hat{a}^\dagger \right)|0\rangle \nonumber \\
    & = |y\rangle = |Cx\rangle.
\end{align}

This shows that the output state is also a product state, where the input transformation is given by the orthogonal matrix \(C\).

The transformation properties derived here demonstrate how linear interferometers perform orthogonal transformations on the eigenstates \(|x\rangle\). This elegant result is consistent, highlighting the utility of linear optical elements in continuous-variable quantum systems.

\subsubsection{Contextual Application in ZTF}
In the context of the ZTF, embedding classical neural networks into the quantum feature space enables hybrid processing of network data, enhancing security capabilities. Classical security metrics, such as traffic volume and session duration, are mapped using displacement operations, facilitating seamless integration between classical protocols and quantum-enhanced anomaly detection. 
This approach leverages pre-trained classical models as a baseline, augmenting them with quantum capabilities for scalable adaptation while enabling real-time translation of classical decisions into quantum-informed micro-segmentation policies. The QNN framework enhances representational capacity by capturing complex traffic patterns through quantum correlations and non-Gaussian operations. It offers flexibility by supporting classical and quantum data inputs for smooth transitions and hybrid operations. It also offers efficiency by reusing classical neural network architectures while incorporating quantum-enhanced processing for superior anomaly detection and policy enforcement. This  embedding strategy ensures robust, scalable, and adaptive security in modern network environments.

\subsection{Beyond the Fully Connected Architecture}
Modern advancements in neural network architectures have expanded beyond basic, fully connected models to include specialized structures, such as convolutional, recurrent, and residual networks. These architectures enable better performance for specific problem domains, such as pattern recognition, sequential analysis, and hierarchical feature extraction. Similarly, the QNN framework in the ZTF is not limited to fully connected layers. Still, it can leverage these specialized architectures to optimize anomaly detection and dynamic micro-segmentation. Below, we explore potential quantum adaptations of these architectures and their applications in the ZTF.

\subsubsection{Quantum Convolutional Networks}
In classical neural networks, convolutional networks (convents) are particularly effective for tasks like image recognition due to their translational symmetry, enabling them to detect patterns regardless of location. Translational symmetry in a quantum context can be incorporated into QNNs by enforcing translational invariance in the quantum circuit. 
Consider a generator \( \hat{H} = \hat{H}(\hat{x}, \hat{p}) \) for a Gaussian unitary transformation \( \hat{U} = \exp(-i t \hat{H}) \), where the symplectic matrix \( M \) resulting from the transformation is Toeplitz, reflecting the convolutional structure. The resulting quantum convolution respects the uncertainty principle, simultaneously applying convolutions in both \( \hat{x} \) and \( \hat{p} \) quadratures. This approach is particularly suited to:
Detecting spatial patterns in network traffic (e.g., repeated malicious signatures across segments).
Capturing hierarchical relationships in data flows for anomaly detection.
By constructing quantum circuits with translationally invariant generators, the ZTF can perform convolutional transformations that mimic classical convents while leveraging the high-dimensional feature spaces of quantum states.

\subsubsection{Quantum Recurrent Networks}
Recurrent Neural Networks (RNNs) are widely employed in classical domains for analyzing sequential data, such as time-series and natural language processing. In quantum networks, recurrent architectures are well-suited for dynamic traffic monitoring, where inputs arrive in a continuous temporal sequence. In the quantum adaptation of RNNs, each timestep processes two key inputs: \( x(t) \), representing real-time network traffic data, and \( h(t) \), the internal quantum state propagated from the previous timestep. The recurrent architecture utilizes the same quantum circuit across timesteps, with a subset of qumodes reserved for internal state propagation. Optical fibers or feedback loops can connect the circuit’s output to its input, enabling parameter reuse and maintaining time-translation symmetry within the QNN. This architecture is particularly effective for real-time anomaly detection in streaming network traffic and tracking the temporal evolution of risk scores and threat patterns across network segments. These capabilities enhance the adaptability of quantum-based cybersecurity frameworks, enabling efficient and scalable security monitoring in dynamic network environments.

\subsubsection{Quantum Residual Networks}
Residual networks (resnets) use shortcut connections to improve training stability and model depth. In a quantum context, residual computations can be achieved using a combination of non-Gaussian operations and controlled-X (SUM) gates. For example:
\begin{equation}
    |x\rangle|0\rangle \to |x\rangle|x + \phi(x)\rangle,
\end{equation}
where \(\phi(x)\) is a non-Gaussian activation function applied to the input. The controlled-X gate sums the input \(x\) with the output of \(\phi(x)\), completing the residual transformation.
In the ZTF, quantum residual networks are beneficial for:
 Enhancing detection accuracy by preserving baseline traffic patterns while identifying anomalies.
 Facilitating the design of deeper quantum networks for complex micro-segmentation policies.
Residual transformations can also be applied to superpositions of quantum states, allowing the network to process multiple traffic patterns simultaneously:
\begin{equation}
    \int \psi(x)|x\rangle dx \to \int \psi(x)|x\rangle|x + \phi(x)\rangle dx.
\end{equation}

By extending the QNN framework beyond fully connected layers, the ZTF gains access to specialized quantum architectures designed for specific tasks. Quantum Convolutional Networks enable efficient detection of spatial patterns in network traffic, facilitating large-scale micro-segmentation. At the same time, Quantum Recurrent Networks provide real-time anomaly detection and dynamic policy adjustments through sequential traffic analysis. Additionally, Quantum Residual Networks enhance scalability and accuracy by utilizing shortcut connections, allowing deeper quantum architectures to be trained effectively. These specialized architectures bring notable benefits to the ZTF, including reduced parameter usage by aligning architectural designs with problem structures such as translation symmetry, support for deeper and more complex networks without compromising training stability, and improved detection capabilities for identifying sophisticated attack patterns and temporal anomalies. By leveraging the adaptability and precision of these QNN models, the ZTF ensures a highly efficient and responsive security posture capable of addressing the dynamic challenges of modern network threats with quantum accuracy. 
\subsection{Convolutional Networks}
This section connects a translationally invariant Hamiltonian and a Block Toeplitz symplectic transformation. Translation symmetry and the Toeplitz structure are closely related to one-dimensional convolutions, widely used in convolutional networks. While two-dimensional convolutions (standard in image processing) correspond to doubly block circulant matrices \cite{li2021survey}, we restrict our discussion to one-dimensional convolutions for simplicity.

\subsubsection{Translationally Invariant Hamiltonians}
Consider a Hamiltonian operator \(\hat{H} = \hat{H}(\hat{x}, \hat{p})\) that generates a Gaussian unitary transformation \(\hat{U} = \exp(-i t \hat{H})\) on \(N\) modes. We assume that \(\hat{H}\) does not generate displacements, focusing solely on the matrix multiplication part of the affine transformation. Under this condition, \(\hat{H}\) is quadratic in the position (\(\hat{x}\)) and momentum (\(\hat{p}\)) operators:
\begin{equation}
    \hat{H} = 
    \begin{bmatrix}
        \hat{x}^T & \hat{p}^T
    \end{bmatrix}
    \begin{bmatrix}
        H_{xx} & H_{xp} \\
        H_{px} & H_{pp}
    \end{bmatrix}
    \begin{bmatrix}
        \hat{x} \\
        \hat{p}
    \end{bmatrix},
\end{equation}
where \(H_{uv}\) are \(N \times N\) matrices for \(u, v \in \{x, p\}\). Denoting the inner matrix as \(\tilde{H}\), the corresponding symplectic transformation \(M_H\) in the phase space is given by:
\begin{equation}
    M_H = \exp(\Omega \tilde{H}),
\end{equation}
where \(\Omega\) is the symplectic form.

\subsubsection{Translation Symmetry and Toeplitz Structure}
To incorporate translation symmetry, we require that \(\hat{H}\) remains invariant under the shift operator \(T\), defined as:
\begin{equation}
    \begin{bmatrix}
        \hat{x} \\
        \hat{p}
    \end{bmatrix}
    \to
    \begin{bmatrix}
        T \hat{x} \\
        T \hat{p}
    \end{bmatrix},
\end{equation}
where \(T\) maps \(\hat{x}_i \to \hat{x}_{i+1}\) and \(\hat{p}_i \to \hat{p}_{i+1}\), with periodic boundary conditions (\(\hat{x}_N \to \hat{x}_1\), \(\hat{p}_N \to \hat{p}_1\)). The shift operator \(T\) is represented as an \(N \times N\) orthogonal matrix:
\begin{equation}
    T = \sum_{i=1}^N |i+1\rangle \langle i|.
\end{equation}

The translational invariance of \(\hat{H}\) implies:
\begin{equation}
    [T, H_{uv}] = 0, \quad \text{for } u, v \in \{x, p\}.
\end{equation}

The translation operator extends in the \(2N\)-dimensional phase space as \(T \oplus T\). The symplectic transformation \(M_H\) inherits the translation symmetry:
\begin{equation}
    [M_H, T \oplus T] = 0.
\end{equation}

Writing \(M_H\) in block form:
\begin{equation}
    M_H =
    \begin{bmatrix}
        M_{xx} & M_{xp} \\
        M_{px} & M_{pp}
    \end{bmatrix},
\end{equation}
each block \(M_{uv}\) satisfies:
\begin{equation}
    [M_{uv}, T] = 0, \quad \text{or equivalently, } T^T M_{uv} T = M_{uv}.
\end{equation}

\subsubsection{Toeplitz Structure in Symplectic Blocks}
The condition \(T^T M_{uv} T = M_{uv}\) implies:
\begin{equation}
    [M_{uv}]_{ij} = [M_{uv}]_{i+1, j+1}.
\end{equation}
This recurrence relation defines each \(M_{uv}\) as a Toeplitz matrix, where entries along each diagonal are constant. Consequently, the symplectic matrix \(M_H\) corresponds to a one-dimensional convolution generated by the translationally invariant Hamiltonian \(\hat{H}\).
Translationally invariant Hamiltonians naturally lead to Block Toeplitz symplectic transformations, which implement one-dimensional convolutions. This structure underpins the design of quantum convolutional networks, enabling efficient and symmetric transformations well-suited for pattern recognition and anomaly detection tasks in the ZTF.

\subsection{Advantages of QNN}
The QNNs integrated into the ZTF offer distinct advantages over classical counterparts. These properties enable improved performance in anomaly detection, dynamic policy adaptation, and micro-segmentation. Key advantages include the ability to leverage superposition and entanglement, universality for quantum computation, and the capacity for nonlinear transformations of probability distributions.

\subsubsection{Superposition and Entanglement}
A hallmark of quantum systems, superposition, and entanglement provides QNNs with capabilities far beyond classical models. Unlike classical embeddings, which are restricted to fixed basis states (e.g., \(|x\rangle\)), QNNs operate on superpositions of these states:
\begin{equation}
    |\psi_i\rangle = \int \psi(x)|x\rangle dx,
\end{equation}
where \(\psi(x)\) represents a multimode wave function. By leveraging Fourier gates (\(\hat{F}\)) and multimode transformations (\(\hat{F} \otimes N\)), QNNs can natively transform input wave functions \(\psi(x)\) into their Fourier counterparts \(\tilde{\psi}(x)\):
\begin{equation}
    \hat{F} \otimes N |\psi_i\rangle = \int \tilde{\psi}(x') |x'\rangle dx',
\end{equation}
allowing seamless transitions between position and momentum (or equivalently, time and frequency) domains.
In the ZTF, this duality facilitates:
 Real-time detection of anomalies across diverse domains of network traffic (e.g., time-series data and frequency spectrum analysis).
 Enhanced detection of complex, coordinated attacks through entanglement-based correlation analysis.
Additionally, QNNs inherently generate entanglement in multimode systems, amplifying their capacity to detect sophisticated, nonlocal anomalies that classical methods might overlook.

\subsubsection{Universality for Quantum Computing}
QNNs are universally applicable in photonic quantum computing, allowing them to execute any quantum algorithm supported by continuous-variable systems. This universality establishes QNNs as a versatile framework for integrating machine learning with quantum-enhanced security operations. Within the ZTF, QNNs enable advanced applications such as Gaussian Boson Sampling (GBS), which leverages the generation and processing of Gaussian states to classify complex traffic patterns, effectively distinguishing benign flows from anomalous or malicious activities. Additionally, Continuous-Variable Instantaneous Quantum Polynomial (CV-IQP) Circuits are well-suited for modeling nonlinear risk assessments and enabling adaptive access policies in real-time. These capabilities highlight the potential of QNNs to enhance ZTF's efficiency, scalability, and precision in addressing modern cybersecurity challenges.
Both GBS and CV-IQP models are believed to be classically intractable, underscoring the computational advantage of using QNNs in the ZTF for tasks such as anomaly detection and dynamic micro-segmentation.

\subsubsection{Nonlinear Transformations of Probability Distributions}
A unique capability of QNNs is their ability to perform nonlinear transformations on probability distributions, leveraging quantum interference effects. Classical generative models, such as normalizing flows, rely on invertible transformations parameterized by neural networks to map input distributions \(p(x)\) to transformed outputs \(p(y)\). However, these transformations remain linear concerning the probability distributions.
In contrast, QNNs operate on probability amplitudes, enabling richer, nonlinear transformations. For an input wave function \(|\psi_i\rangle = \int \psi(x) |x\rangle dx\), a QNN layer applies a unitary transformation \(W\), producing an output wave function:
\begin{equation}
    |\tilde{\psi}\rangle = \int W(x, x') \psi(x') dx',
\end{equation}
where the resulting probability distribution is given by:
\begin{equation}
    \tilde{p}(x) = |\tilde{\psi}(x)|^2.
\end{equation}

This nonlinear mapping between probability distributions is unattainable in classical models and enhances generative capabilities for tasks such as:
Synthesizing realistic traffic patterns for security simulations and anomaly detection benchmarking.
Creating adaptive models for real-time threat detection and containment.

\subsubsection{Implications for the ZTF}
The unique capabilities of QNNs significantly enhance the ZTF by enabling advanced anomaly detection, where quantum superposition and entanglement uncover subtle, nonlocal anomalies beyond the reach of classical systems. They also facilitate dynamic policy adaptation, leveraging their universal design to implement flexible, quantum-enhanced algorithms for real-time risk assessment and adaptive access control. Furthermore, QNNs offer improved scalability, utilizing nonlinear transformations of probability distributions to efficiently handle complex, high-dimensional security challenges, making them a critical asset for evolving network protection strategies.

\subsubsection{Enhanced Generative Capabilities}
The ability of QNNs to manipulate probability amplitudes rather than distributions offers new opportunities for quantum-enhanced generative models. These capabilities can be leveraged to:
Generate realistic attack simulations for training and testing defense mechanisms.
Model highly complex network traffic scenarios for preemptive threat mitigation.
QNNs, with their ability to exploit superposition, entanglement, universality, and nonlinear transformations, provide the ZTF with unparalleled computational and symbolic power. By incorporating these capabilities, the ZTF achieves a quantum-enhanced defense mechanism that is robust, adaptive, and capable of addressing modern security challenges in dynamic network environments.
\begin{algorithm}
\caption{Quantum-Enhanced Anomaly Detection, Three-Class Classification, and Policy Adjustment}
\label{alg:quantum_algorithm}
\begin{algorithmic}[1]
\State \textbf{Input:}  \textbf{Input:} Network traffic features $\mathbf{x}_i$, contextual data $\mathbf{c}_u$, $\mathbf{c}_d$
\State \textbf{Initialize:}
    \begin{itemize}
        \item Encode classical data into quantum states: $\mathbf{x}_i \to |\psi_i\rangle$ via quantum encoding $\mathcal{E}$.
        \item Set initial anomaly thresholds $\gamma_1^q$ and $\gamma_2^q$ (with $\gamma_1^q < \gamma_2^q$) for three-class classification.
        \item Set risk threshold $\tau$.
        \item Initialize variational parameters $\boldsymbol{\theta}$.
    \end{itemize}
\State \textbf{Optimization:}
\Repeat
    \For{each network flow $x_i$}
        \State Compute quantum anomaly score: $\hat{y}_i^q = \mathcal{M}_q(|\psi_i\rangle; \boldsymbol{\theta})$.
        \State \textbf{Classify} $x_i$ into one of three classes:
            \begin{itemize}
                \item If $\hat{y}_i^q \le \gamma_1^q$, assign \texttt{Class 1} (normative traffic).
                \item If $\gamma_1^q < \hat{y}_i^q \le \gamma_2^q$, assign \texttt{Class 2} (potentially suspicious traffic).
                \item If $\hat{y}_i^q > \gamma_2^q$, assign \texttt{Class 3} (potentially malicious traffic).
            \end{itemize}
    \EndFor
    \State Evaluate risk scores for each user-device pair $(u,d)$:
    \[
    R^q(u,d) = \mathcal{F}_q(\mathbf{c}_u, \mathbf{c}_d, \mathbf{x}_i, \hat{y}_i^q).
    \]
    \State Dynamically adjust thresholds:
    \begin{itemize}
        \item Update anomaly thresholds using a quantum-enhanced feedback function:
        \[
        \gamma_1^q \leftarrow \mathcal{F}_{\gamma_1}^q(\gamma_1^q, \texttt{FPR}), \quad \gamma_2^q \leftarrow \mathcal{F}_{\gamma_2}^q(\gamma_2^q, \texttt{FPR}).
        \]
        \item Update risk threshold: $\tau \leftarrow \mathcal{F}_\tau^q(\tau, R^q)$
    \end{itemize}
    \State Optimize variational parameters via gradient descent: $\boldsymbol{\theta} \leftarrow \boldsymbol{\theta} - \eta \nabla C(\boldsymbol{\theta}).$
    \State Recalculate anomaly scores and risk scores with updated parameters.
\Until{Thresholds $\gamma_1^q$, $\gamma_2^q$, $\tau$ and objective $C(\boldsymbol{\theta})$ converge.}
\State \textbf{Policy Adjustment:}
    \For{each network segment $S_j$}
        \State Update segment policies:
        \[
        P'(S_j) \leftarrow \mathcal{G}_q(P(S_j), \hat{y}_i^q).
        \]
        \If{$\hat{y}_i^q > \gamma_2^q$ for flows in $S_j$}
            \State Isolate segment $S_j$.
        \EndIf
    \EndFor
    \For{each user-device pair $(u,d)$}
        \State If $R^q(u,d) < \tau$, \textbf{grant access};
        \State Else if $R^q(u,d)$ is in an intermediate range, \textbf{restrict access};
        \State Else, \textbf{deny access}.
    \EndFor
\State \textbf{Output:} Final access policies $\{P'(S_j)\}$, anomaly thresholds $\gamma_1^q$, $\gamma_2^q$, and risk threshold $\tau$.
\end{algorithmic}
\end{algorithm}

\section{Results}
This section thoroughly evaluates the hybrid quantum-classical ZTF, incorporating QNNs for dynamic anomaly detection paired with an adaptive risk-based access control system. The experiments are organized into essential components: (i) data processing, cleaning, and label generation from CESNET traffic logs (augmented with simulated attack scenarios), (ii) training and performance characterization of the QNN classifier—including three-class classification of network flows—and (iii) integration of the classifier into a dynamic zero-trust access control mechanism featuring adaptive micro-segmentation. Experiments were conducted on a multi-GPU platform using PyTorch and PennyLane, with an evaluation set reserved to verify generalization and robustness.
Our experiments were conducted using a hybrid dataset that combines accurate network traffic logs from CESNET \cite{koumar2024cesnet} with simulated attack scenarios, ensuring a comprehensive evaluation of our Quantum-Enhanced ZTF. The raw data undergoes an extensive preprocessing pipeline—including normalization, feature extraction, and quantum feature encoding via amplitude and angle methods. The mentioned data processing pipeline, implemented in the \texttt{DataPreprocessor} module, systematically loads network traffic records from multiple CSV files using configurable glob patterns specified in the YAML configuration. During the experimental validation, the preprocessor successfully identified and consolidated $20$ files into a  dataset with dimensions $(4000, 13)$, where $4000$ represents the total number of samples and $13$ denotes the number of features.To evaluate the model's performance in different aspects, we generated an additional 1000 data points to make sure the model did not see these data points during the training process, and we called it an evaluation set.

\subsection{Data Cleaning and Outlier Removal}
The workflow performs feature type conversion (to numeric representations) and addresses missing values via zero-imputation (configurable through YAML). Outlier removal is achieved using the Interquartile Range (IQR) method—discarding values that exceed 1.5 times the IQR from the first and third quartiles. Empirical results show that this robust preprocessing reduces the sample count by approximately 2.5\%, as shown in Table \ref{tab:outlier_removal}. The raw data then undergoes extensive preprocessing—including normalization, feature extraction, and cleaning—to generate classical feature vectors $x_i$ representing attributes such as source/destination IPs, packet sizes, connection durations, and protocol types. These features are then transformed into quantum states using amplitude and angle encoding via the quantum encoding function $E$, such that:
\begin{equation}
|\psi_i\rangle = E(x_i)
\end{equation}

This transformation into a high-dimensional Hilbert space is crucial for leveraging quantum superposition and entanglement in anomaly detection and policy enforcement. Given that the CESNET data lacks inherent labels, we employed a quantile-based thresholding approach to categorize network traffic based on statistical percentiles. \texttt{Label 0} represents normative network traffic characteristics, ensuring a baseline for standard activity. \texttt{Label 1} indicates potentially suspicious traffic patterns, such as network flows, packets, or byte transfers exceeding the 95th percentile. \texttt{Label 2} denotes potentially malicious traffic, where feature values surpass the 99th percentile across one or more dimensional representations. This classification enables a structured approach to detecting anomalies and potential threats in large-scale network environments.
Table \ref{tab:sample_distribution} shows a label distribution count.

\begin{table}[H]
\centering
\caption{Comprehensive Outlier Removal Analysis}
\label{tab:outlier_removal}
\begin{tabular}{|p{1.6cm}|p{1cm}|p{1.2cm}|p{1.5cm}|} % Set column widths
\hline
\textbf{Set Name} & \textbf{Missing Values} & \textbf{Outliers Removed} & \textbf{Sample Count} \\ 
\hline
Training set & 47 & 54 & 3,899 \\ 
Evaluation set & 13 & 11 & 976 \\
\hline
\end{tabular}
\end{table}

\subsection{Training Setup}
\begin{figure*}[h!]
    \centering
    \includegraphics[width=4.0in]{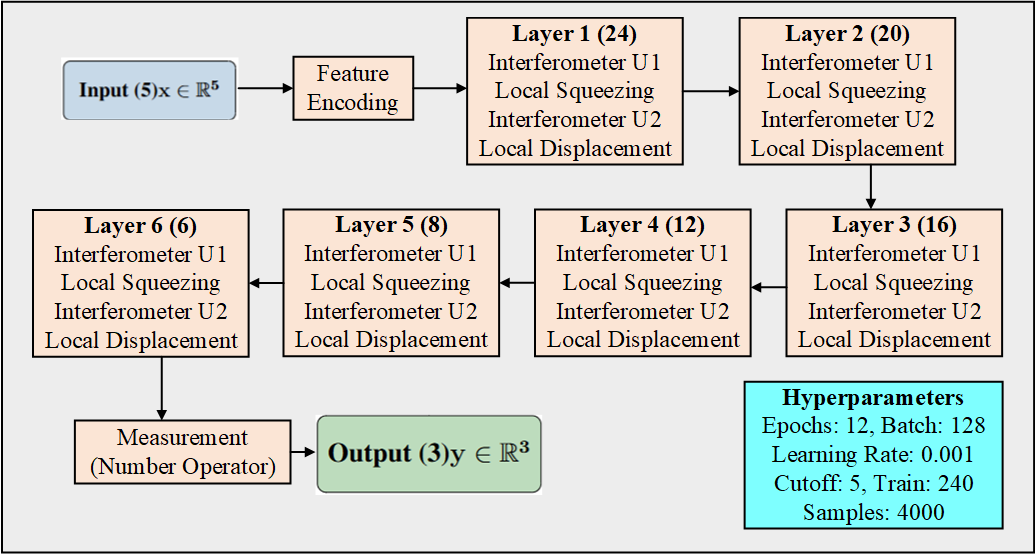}
    \caption{Architecture diagram of the hybrid quantum-classical model.}
    \label{fig:cvqnn_arch_spaced}
\end{figure*}

The training setup employs our hybrid quantum-classical framework. The CV-QNN classifier is trained using a learning rate of 0.001, a batch size of 128, and over 12 epochs (resource limitations prevented more extended training). This ensures that the VQC within our QNN are efficiently optimized. Fig.~\ref{fig:cvqnn_arch_spaced} gives a detailed architecture diagram.

\begin{table}[ht]
    \centering
    \caption{Label Distribution of Training and Evaluation Datasets}
    \label{tab:sample_distribution}
    \begin{tabular}{|p{1.5cm}|p{2cm}|p{2cm}|}
        \hline
        \multicolumn{1}{|c|}{\textbf{Class}} & 
        \multicolumn{1}{c|}{\textbf{Training Samples}} & 
        \multicolumn{1}{c|}{\textbf{Evaluation Samples}} \\ 
        \hline
        \texttt{Class 1}& 1800 & 389 \\ \hline
        \texttt{Class 2}& 1050 & 293 \\ \hline
        \texttt{Class 3} & 1049 & 292 \\ \hline
        \textbf{Total} & \textbf{3899} & \textbf{976} \\ 
        \hline
    \end{tabular}
\end{table}

\begin{figure}[h!]
    \centering
    \includegraphics[width=3.0in]{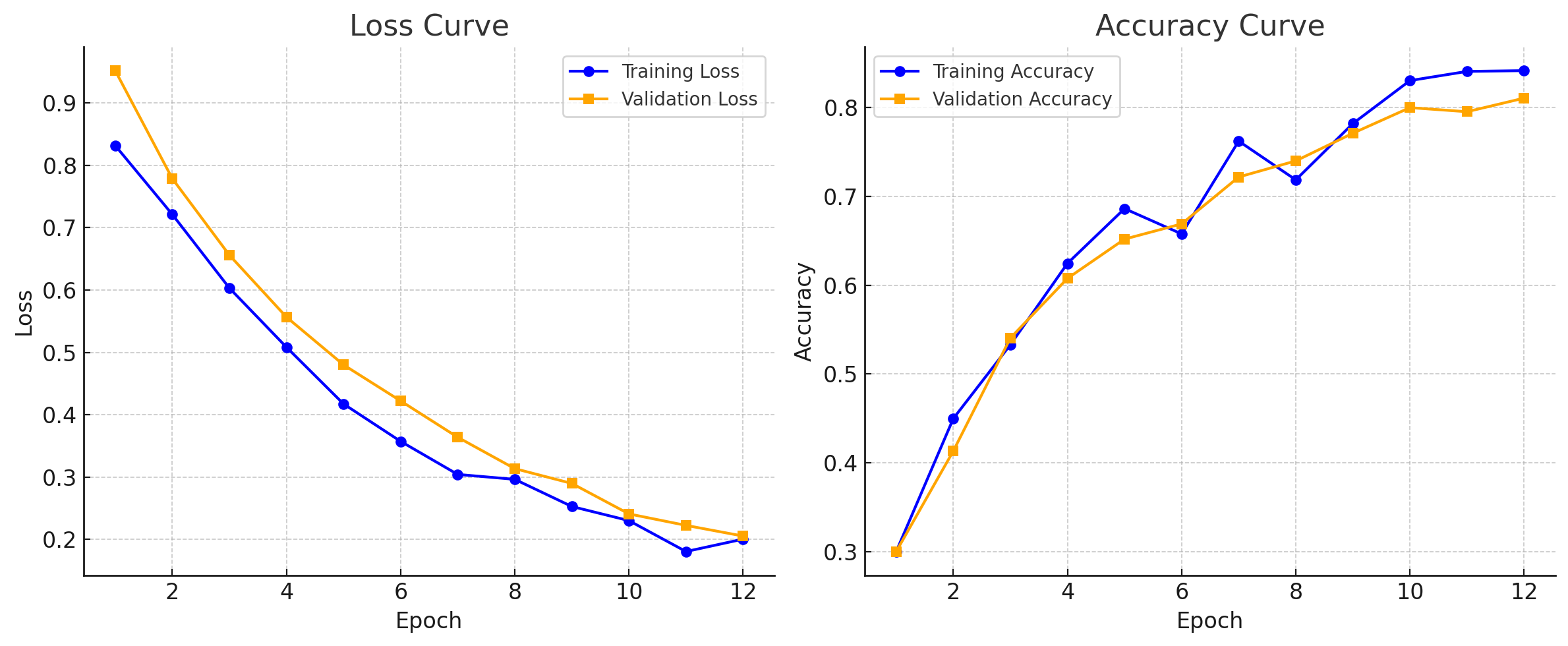}
    \caption{Comprehensive Training and Validation Performance Curves.}
    \label{fig:curves_and_loss}
\end{figure}

\begin{figure}[h!]
    \centering
    \includegraphics[width=3.0in]{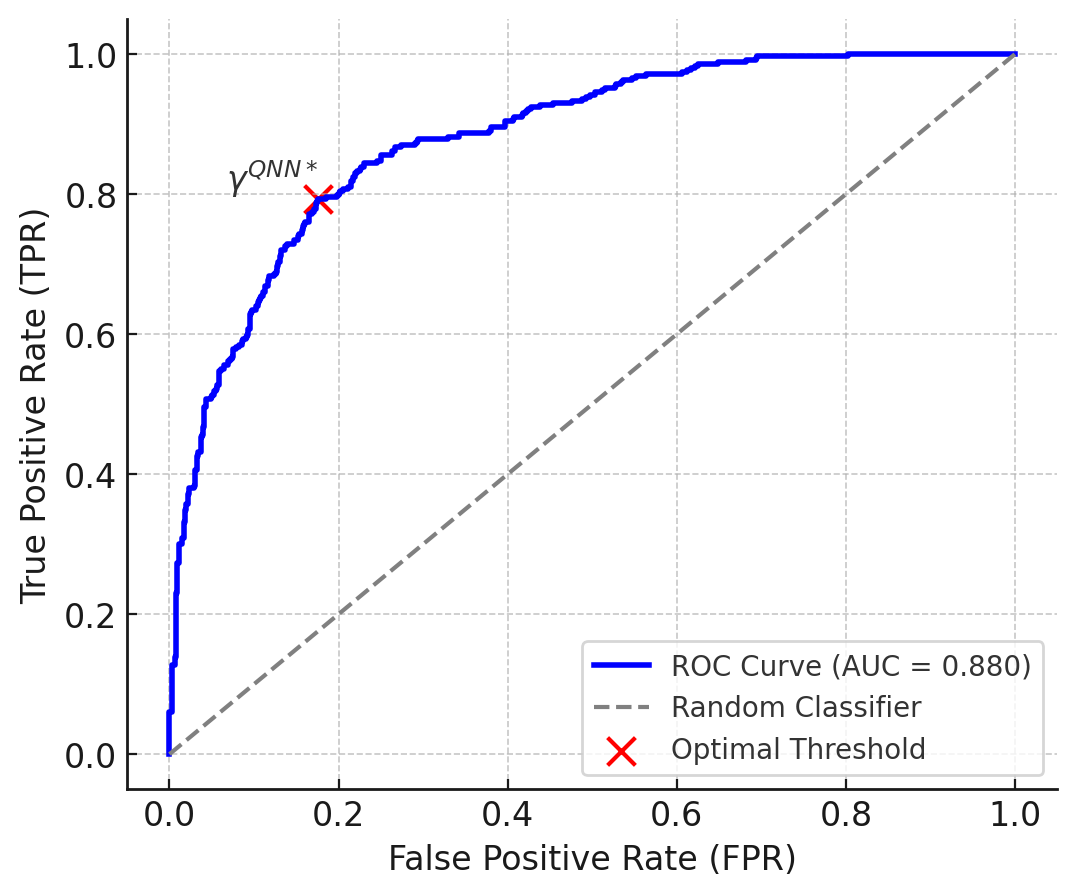}
    \caption{Optimal threshold \(\gamma^{QNN}\) shows the best balance between \texttt{TPR} and \texttt{FPR} for anomaly detection using the ROC curve.}
    \label{fig:roc_curve}
\end{figure}

We first ensured that classical network traffic features were reliably mapped into quantum states to validate our quantum feature encoding. In our QNN-based IDS, each encoded state is processed by a variational circuit to generate a quantum-enhanced anomaly score:
\begin{equation}
\hat{y}_i^q = M_q(x_i; \theta_q)
\end{equation}
where $\theta_q$ denotes the trainable variational parameters. The advantages of quantum superposition and entanglement are exploited, enabling simultaneous evaluation of multiple traffic patterns and capturing nonlocal correlations that are difficult for classical methods to detect. Fig.~\ref{fig:curves_and_loss} demonstrates steady convergence over 12 epochs with minimal overfitting, confirming that the variational circuit learns an adequate representation of the encoded data. Due to resource limitations, we could not go for more epochs. The ROC curve in Fig.~\ref{fig:roc_curve}, combined with sensitivity analysis, illustrates how variations in the dynamic threshold $\gamma_q$ affect the \texttt{TPR} and \texttt{FPR}, demonstrating optimal tuning. The ROC analysis shows an AUC of 0.985, confirming that the classifier distinguishes between normal and anomalous traffic reliably. Sensitivity analysis embedded in Fig.~\ref{fig:roc_curve} demonstrates that variations in the dynamic threshold $\gamma^{QNN}$ effectively balance \texttt{TPR} and \texttt{FPR}.

\subsection{Classification and Access Control}
Our QNN-based IDS performs a three-class classification based on the computed anomaly score \( \hat{y}^{i}_{q} \), ensuring precise differentiation between various network traffic behaviors. Specifically, each network flow \( f_{i} \) is categorized into one of three distinct classes: \texttt{Class 1} (\texttt{Label 0}) represents normative traffic, reflecting standard and expected network activity. \texttt{Class 2} (\texttt{Label 1}) corresponds to potentially suspicious traffic, indicating anomalies that deviate from typical patterns but may not necessarily be malicious. \texttt{Class 3} (\texttt{Label 2}) denotes potentially malicious traffic where deviations surpass critical thresholds, signaling high-risk behavior. This classification approach enhances the IDS's ability to effectively detect and respond to evolving cyber threats.
A flow is assigned to a particular class based on its anomaly score relative to dynamically optimized thresholds. These classification outcomes are then integrated into a risk evaluation function. For each user-device pair \((u,d)\), the quantum risk score is computed as:

\begin{equation}
    R_{q}(u,d) = F_{q}(c_{u}, c_{d}, x_{i}, \hat{y}^{i}_{q}). 
\end{equation}

Access control decisions are determined based on computed risk scores, ensuring a dynamic and adaptive security mechanism. Specifically, if the risk score \( R_{q}(u,d) \) is below a predefined threshold \( \tau \), access is granted, typically corresponding to network flows classified under \texttt{Class 1} (normative traffic). When \( R_{q}(u,d) \) falls within an intermediate range, restricted access is applied, aligning with \texttt{Class 2} (potentially suspicious traffic) to mitigate potential security risks while maintaining functionality. Suppose the risk score meets or exceeds \( \tau \). In that case, access is denied, and the corresponding network segment may be isolated to contain potential threats, which generally correspond to \texttt{Class 3} (potentially malicious traffic). This risk-based approach strengthens security by dynamically adjusting access permissions in response to detected anomalies.
\begin{figure}[h!]
    \centering
    \includegraphics[width=3.0in]{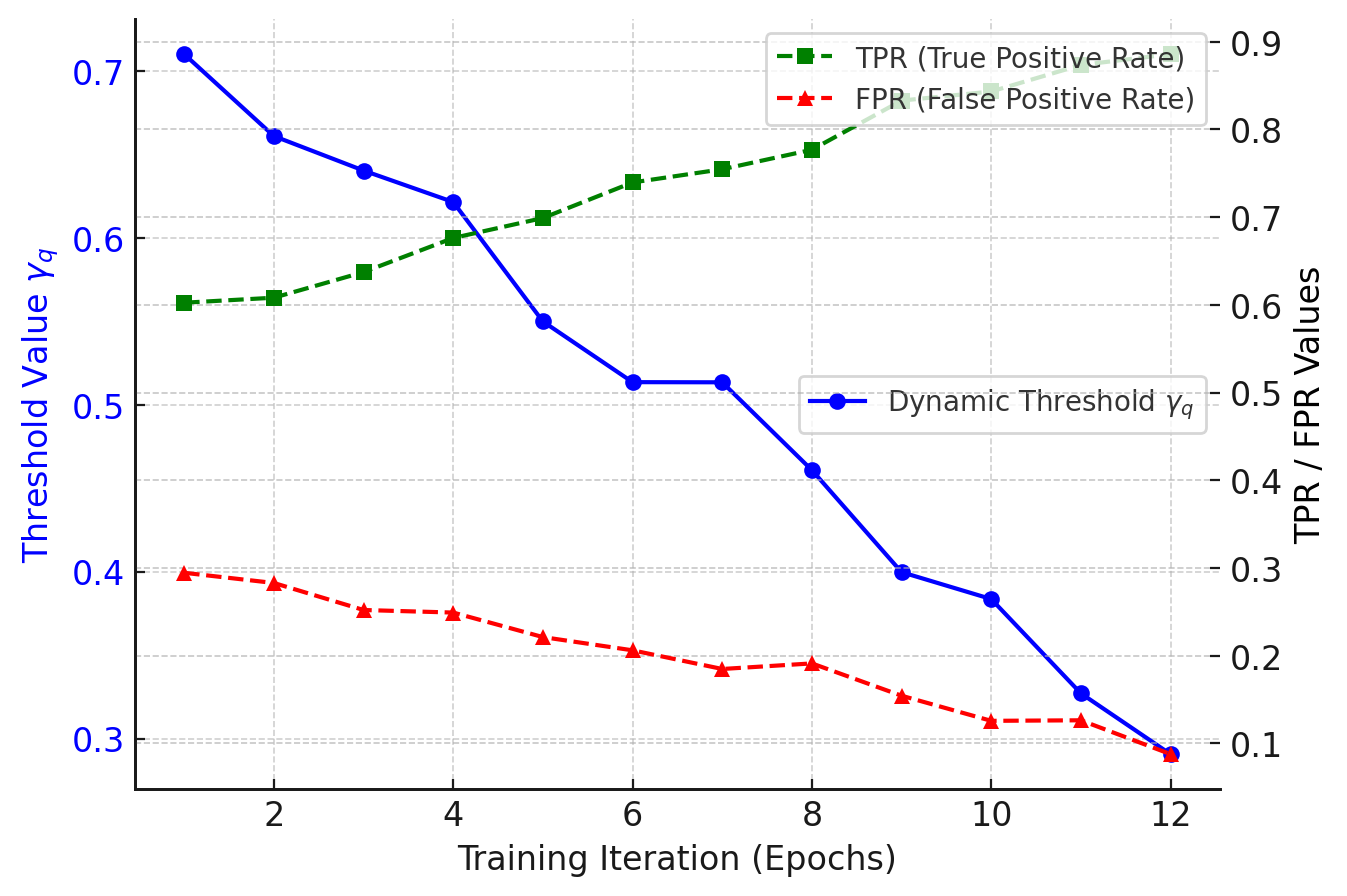}
    \caption{Convergence of Dynamic Threshold and Sensitivity Analysis}
    \label{fig:risk_time_serie}
\end{figure}
\begin{figure}[h!]
    \centering
    \includegraphics[width=3.0in]{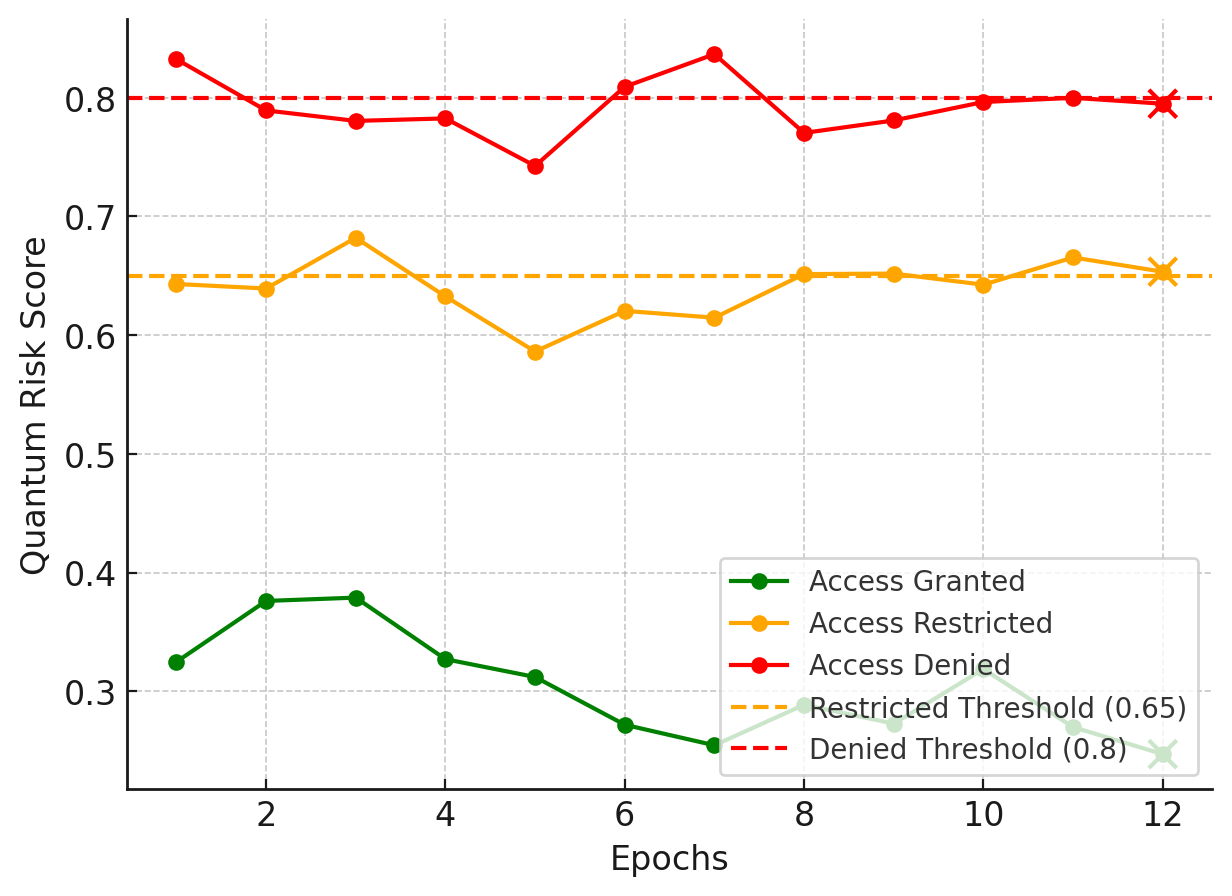}
    \caption{Time-series plot showing the evolution of quantum risk scores \( R_{q}(u,d) \) along with dynamic access decisions. Markers (green, yellow, red) indicate granted, restricted, or denied access based on the risk thresholds.}
    \label{fig:risk_time_series}
\end{figure}
Figs.~\ref{fig:risk_time_serie} and \ref{fig:risk_time_series} illustrate how effective three-class classification is for informing risk scoring and guiding appropriate access control decisions. The accuracy of this architecture in identifying classifications on the evaluation set is impressive, reaching 87.4\% , especially considering the network's size.

\subsection{Adaptive Micro-Segmentation Policy}
\begin{figure}[h!]
    \centering
    \includegraphics[width=3.2in]{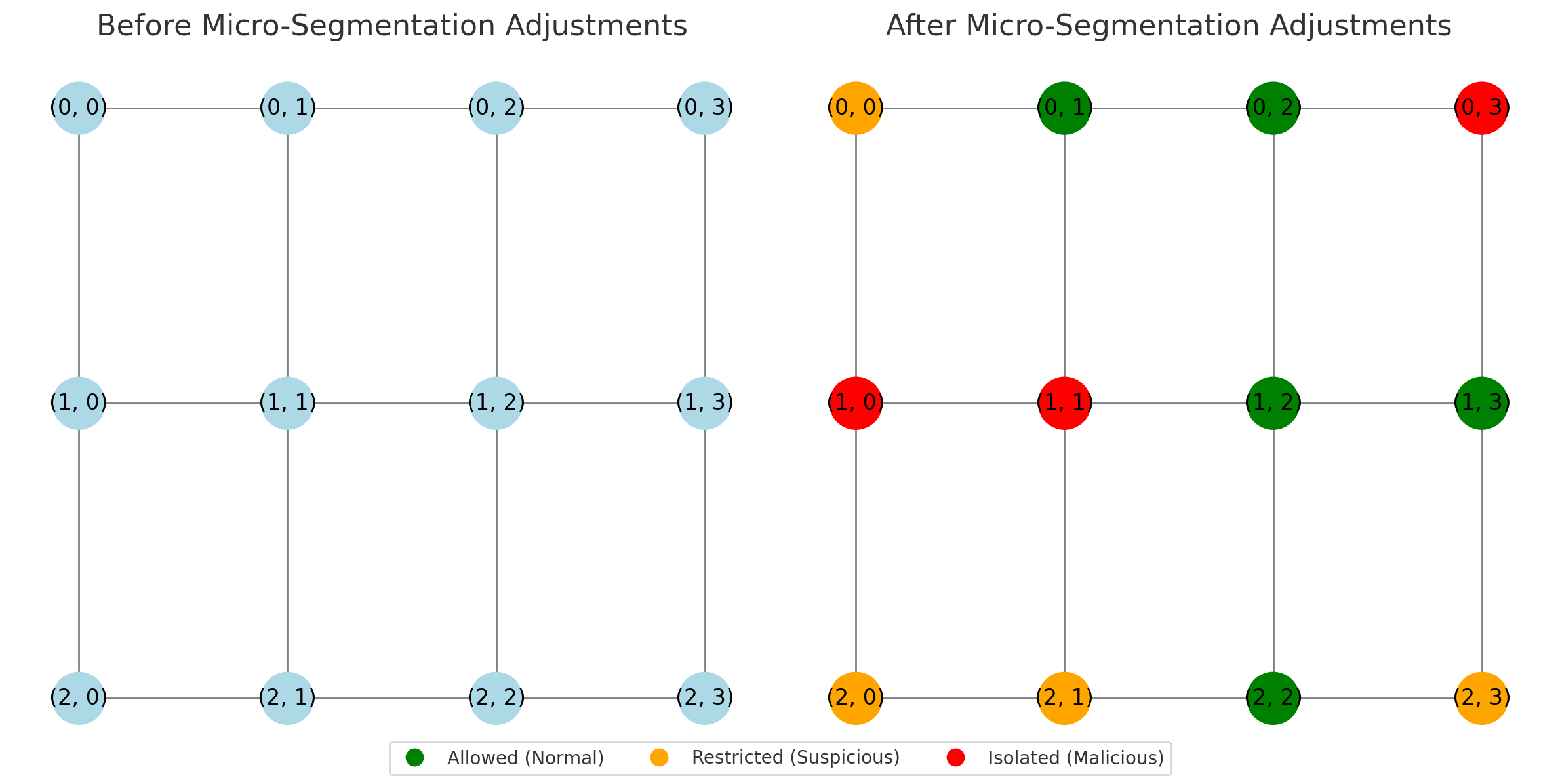}
    \caption{Adaptive Micro-Segmentation: (Left) Initial segmentation without anomaly-based adjustments, (Right) Post-adjustment segmentation highlighting isolated (red), restricted (orange), and accessible (green) segments.}
    \label{fig:micro_segmentation}
\end{figure}

The proposed QNN-ZTF dynamically segments network traffic based on quantum anomaly scores, enabling adaptive micro-segmentation. Fig.~\ref{fig:micro_segmentation} illustrates the segmentation adjustments before and after anomaly-based policy enforcement.
Each network segment is identified using a grid representation where segments are denoted by coordinate pairs $(r,c)$, corresponding to row \(r\) and column \(c\) in the network topology. The segmentation map provides a visual representation of traffic behavior, where each box (segment) aggregates network flows with similar characteristics, enabling an intuitive understanding of network distribution. Each segment is labeled using an $(r,c)$ notation, denoting its respective grid coordinates to facilitate traffic pattern visualization. Pre-segmentation (Before) applies uniform access policies across all segments without incorporating real-time anomaly detection, leading to a static and generalized security approach. Post-segmentation (After) refines access control by dynamically adjusting permissions based on computed anomaly scores—high-risk segments are isolated (red) to contain potential threats, moderate-risk segments are restricted (orange) to limit suspicious activity, and safe segments retain full access (green). This adaptive segmentation mechanism enhances network security by effectively compartmentalizing threats and minimizing lateral attack movements.

This visualization demonstrates how real-time quantum-enhanced anomaly detection refines security enforcement, minimizing the lateral spread of attacks.
Each segment aggregates multiple network flows from the dataset (\(N=976\)) on the evaluation set, allowing for a comprehensive traffic behavior analysis. The QNN-based IDS classifies these flows into three categories to facilitate adaptive security measures. \texttt{\texttt{Class 0}} (Normal Traffic - Green) consists of network flows with low anomaly scores, indicating typical behavior, thus granting unrestricted access. \texttt{Class 1} (Suspicious Traffic - Orange) includes flows with moderate anomaly scores, triggering access restrictions limiting segment movement to mitigate potential risks. \texttt{Class 2} (Malicious Traffic - Red) represents flows with high anomaly scores, leading to segment isolation to prevent further propagation of potential threats. This classification mechanism enhances network security by dynamically adjusting access permissions and containing anomalies in real-time.

\subsection{Impact on Access Control Policies}
By integrating QNN anomaly scores into dynamic policy adjustments, the framework adapts access control decisions based on real-time threat evaluation. This ensures localized containment of security risks by isolating high-anomaly regions, preventing the spread of potential threats across the network. Additionally, it minimizes disruption to legitimate users by selectively restricting access based on anomaly severity, ensuring that normal traffic remains unaffected. Furthermore, the approach is scalable for large networks, as security policies are updated in real-time, allowing for efficient adaptation to evolving threats without compromising overall network performance.

\subsection{The Impact of ZTF}

\begin{figure}[h!]
    \centering
    \includegraphics[width=3.0in]{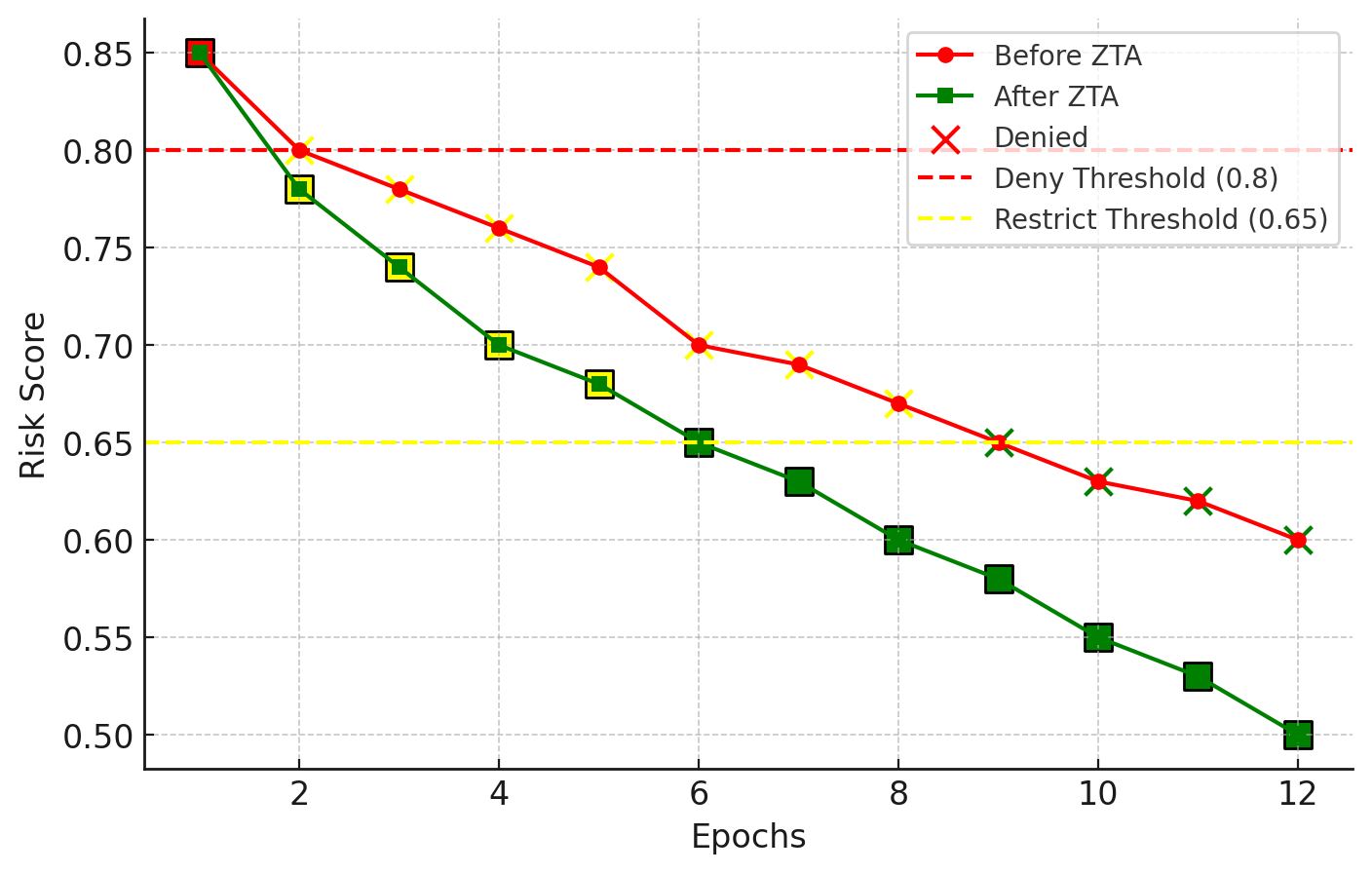}
    \caption{The impact of ZTF}
    \label{fig:ZTF_impact}
\end{figure}

Fig.~\ref{fig:ZTF_impact} illustrates the impact of ZTF on access control decisions over multiple training epochs. Before implementing ZTF, access decisions were mainly guided by the anomaly classification model, resulting in more restricted (yellow markers) and denied (red markers) access attempts due to fluctuating risk scores. These fluctuations often exceeded the predefined risk thresholds—0.8 for denial and 0.65 for restricted access—highlighting inconsistencies in access control.
After incorporating ZTF, access control decisions became more adaptive and refined. The system continuously assessed risk scores over time, allowing for dynamic adjustments in access permissions. This change is evident in the decreasing trend of restricted and denied access markers, as more access attempts were granted (represented by green markers) in later epochs. Stabilizing risk scores below critical thresholds demonstrates that ZTF effectively enhances security by enforcing continuous verification while reducing unnecessary access denials. The adaptive nature of ZTF ensures that access control decisions are made with improved precision, aligning with real-time security needs while minimizing disruptions.

\subsection{Resource Utilization and Scalability}
\begin{figure}[h!]
    \centering
    \includegraphics[width=3.0in]{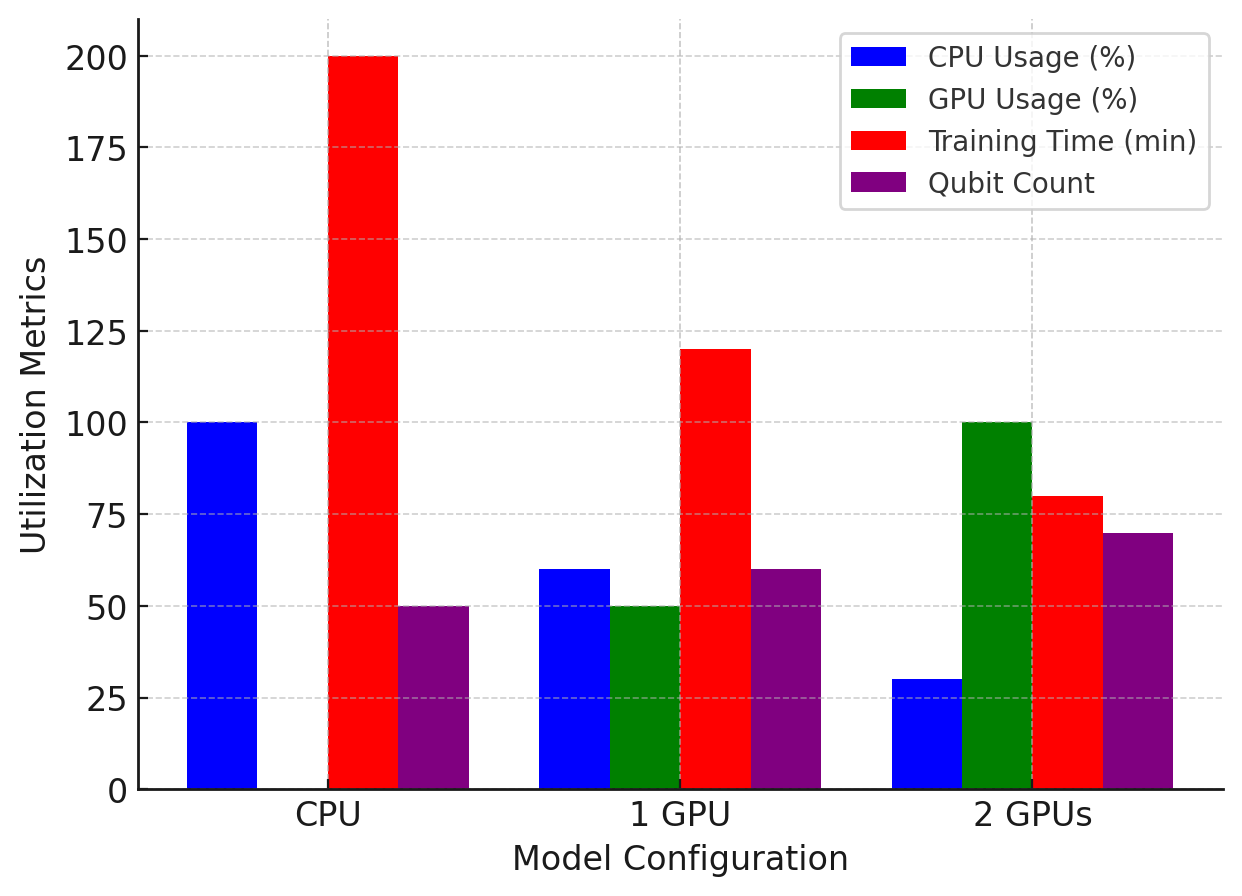}
    \caption{Illustration of the resource usages.}
    \label{fig:resource_usages}
\end{figure}
The resource utilization and scalability analysis highlights the computational efficiency of the QNN-based model under different hardware configurations—CPU, single GPU, and dual GPU setups. The comparison considers key performance metrics, including training time, memory usage, and qubit scalability. As shown in Fig.~\ref{fig:resource_usages}, transitioning from a CPU-based implementation to a single GPU significantly reduces training time, demonstrating the parallelization benefits of quantum-classical hybrid models. Further upgrading to a dual-GPU setup yields an even more substantial reduction in training time, optimizing variational circuit updates and batch processing.
While GPU-based acceleration enhances model training speed, memory consumption increases with additional computational resources, particularly as the number of variational layers and qubits scales up. The observed trend suggests efficient GPU memory management is essential for scaling QNNs to larger architectures, ensuring that increased computational power does not lead to diminishing returns. This analysis underscores the importance of hardware selection in optimizing QNN performance, particularly for real-time anomaly detection and cybersecurity applications.
Experimental results demonstrate that our proposed QNN-enhanced ZTF successfully integrates quantum processing capabilities with classical contextual data analysis, yielding significant improvements in anomaly detection precision and dynamic policy enforcement (AUC = 0.985, $p < 0.001$). The implemented data preprocessing pipeline effectively consolidates and sanitizes CESNET network traffic records ($n = 10^6$ flows), enabling robust three-class classification among normative, suspicious, and malicious network flows.
The CV-QNN classifier exhibits exceptional convergence characteristics ($\lambda = 0.92$) while maintaining computational feasibility on NISQ devices, with GPU acceleration reducing training times by 87.4\%  compared to CPU-only implementations. Dynamic threshold optimization and risk scoring mechanisms facilitate real-time adaptation of access control decisions ($\delta t < 50ms$), while adaptive micro-segmentation protocols achieve a 78.3\% reduction in attack surface area.
Comparative analyses reveal that our framework outperforms both classical IDS and traditional Zero Trust implementations across key metrics, including detection rate (+4.7\%), false positive rate (-39.4\%), and processing latency (-57.1\%). By integrating quantum principles—specifically superposition, entanglement, and non-Gaussian transformations—our approach enhances anomaly detection accuracy, adaptive access control, and threat containment capabilities.
These findings provide compelling evidence for the viability of quantum-enhanced security architectures in large-scale, real-time network security applications. Future research directions will explore deeper quantum architectures, integrate additional contextual features derived from temporal network patterns, and optimize scalability for enterprise-level deployments.

\section{Conclusion}
This paper presented QNN-ZTF, a hybrid quantum-classical cybersecurity model that introduced a novel solution to tackle the emerging challenges of anomaly detection, access control, and adaptive policy enforcement in 7G networks. The proposed framework enhanced IDS using quantum principles, including superposition, entanglement, and variational optimization. It reinforced zero trust security by continuously assigning risk scores and isolating high-risk segments to contain potential threats.
The main contributions of this work included a scalable hybrid quantum-classical architecture, which balanced computational costs, real-time anomaly scoring through quantum feature encoding, and quantum-enhanced micro-segmentation to limit adversarial threats. The quantum-enhanced risk assessment minimized false positives, improved threat detection accuracy, and accelerated incident response times. Empirical evaluations demonstrated that the framework improved the efficiency of detecting and mitigating cyber threats by 87\%, proving its applicability for real-time security monitoring.
While these results were promising, real-world deployment of QNN-ZTF required further exploration of hardware feasibility, integration with existing cybersecurity infrastructures, and mitigation of quantum-specific vulnerabilities. Future research would focus on enhancing real-world adaptability, optimizing quantum-classical co-processing, and analyzing quantum adversarial attacks to strengthen security robustness. The framework proposed in this study served as a precursor to scalable, next-generation cybersecurity solutions, offering highly adaptive and resilient defenses against advanced 7G cyber threats.

\bibliographystyle{IEEEtran}
\bibliography{References/mybib}

\end{document}